\documentclass[a4paper,11pt]{article}
\usepackage{jcappub} % for details on the use of the package, please see the JINST-author-manual
\usepackage{lineno}
\usepackage{xcolor}

\newcommand{\black}{\textcolor{black}}
\usepackage{subcaption}

\title{\boldmath Numerical simulations of primordial black hole formation via delayed first-order phase transitions}

\author[a,b,c]{Zhuan Ning,}
\emailAdd{ningzhuan17@mails.ucas.ac.cn}

\author[b,d]{Xiang-Xi Zeng,}
\emailAdd{zengxiangxi@itp.ac.cn}

\author[e]{Rong-Gen Cai,}
\emailAdd{caironggen@nbu.edu.cn}

\author[b,f]{Shao-Jiang Wang}
\emailAdd{schwang@itp.ac.cn (corresponding author)}

\affiliation[a]{School of Fundamental Physics and Mathematical Sciences, Hangzhou Institute for Advanced Study (HIAS), University of Chinese Academy of Sciences (UCAS), Hangzhou 310024, China}
\affiliation[b]{Institute of Theoretical Physics, Chinese Academy of Sciences (CAS), Beijing 100190, China}
\affiliation[c]{University of Chinese Academy of Sciences (UCAS), Beijing 100049, China}
\affiliation[d]{School of Physical Sciences, University of Chinese Academy of Sciences (UCAS), Beijing 100049, China}
\affiliation[e]{Institute of Fundamental Physics and Quantum Technology \& School of Physical Science and Technology, Ningbo University, Ningbo, 315211, China}
\affiliation[f]{Asia Pacific Center for Theoretical Physics (APCTP), Pohang 37673, Korea}

\abstract{We perform fully nonlinear, spherically symmetric numerical simulations of superhorizon false-vacuum-domain (FVD) collapse in a coupled gravity-scalar-fluid system to study primordial black hole (PBH) formation during delayed first-order phase transitions (FOPTs). Using adaptive mesh refinement to resolve the bubble wall, we identify three dynamical outcomes: type B (supercritical) PBHs with an interior baby universe and a bifurcating trapping horizon, type A (subcritical) PBHs with an apparent horizon formed by direct wall collapse, and dispersal with no PBH formation. To separate these three cases, we evaluate two commonly used PBH-formation criteria: the time scale ratio $t_\mathrm{H}/t_\mathrm{V}$ (horizon crossing time versus vacuum-energy domination time) and the local density contrast $\delta(t_\mathrm{H})$ at horizon crossing. For the parameter space explored, we find that $t_\mathrm{H}/t_\mathrm{V}$ is a more robust predictor of outcome: type B PBHs form when $t_\mathrm{H}/t_\mathrm{V} \gtrsim 1$ (critical range $\sim 1.1 - 1.6$ in our survey), type A PBHs arise when $t_\mathrm{H}/t_\mathrm{V}$ is below this threshold but remains above a lower bound (typical range $\sim 0.35 - 0.7$), and no-PBH dispersal occurs when $t_\mathrm{H}/t_\mathrm{V}$ falls below this lower bound. When a clear thin-wall FVD boundary exists, $\delta(t_\mathrm{H})$ can correspondingly distinguish different outcomes (roughly $\delta_c(t_\mathrm{H}) \sim 1 - 1.7$ for type B and $\delta_c(t_\mathrm{H}) \sim 0.35 - 0.5$ for type A), but is highly sensitive to wall structure and model details and thus less universal. These results offer new insights into the dynamics of FVD collapse, quantify practical PBH-formation thresholds, and pave the way for precise predictions of PBH abundance from delayed FOPTs.}

\begin{document}
\maketitle
\flushbottom

\section{Introduction} \label{sec:introduction}

Primordial black holes (PBHs) are black holes that formed in the early Universe in a non-stellar way~\cite{Zeldovich:1967lct, Hawking:1971ei, Carr:1974nx, Carr:1975qj, Chapline:1975ojl}. They have garnered considerable attention due to their potential cosmological and astrophysical implications. PBHs provide unique probes of the statistics of small-scale primordial perturbations~\cite{Bugaev:2010bb, Ferrante:2022mui, Inui:2024fgk, Ning:2025ogq, Zeng:2025law, Zeng:2025cer, Farooq:2025qmm, Zeng:2025tno, Ning:2025yvj} and are a natural candidate for cold dark matter without invoking physics beyond the Standard Model of particle physics~\cite{Carr:2016drx, Carr:2020gox, Carr:2020xqk, Green:2020jor, Carr:2026hot}. They have also been proposed as progenitors of some binary black hole mergers observed by LIGO-Virgo-KAGRA~\cite{Bird:2016dcv, Sasaki:2016jop, Sasaki:2018dmp, Yuan:2025avq, DeLuca:2025fln} and as seeds for supermassive black holes~\cite{Bean:2002kx, Duechting:2004dk, Kawasaki:2012kn, Carr:2018rid, Huang:2023chx, Huang:2023mwy, Zhang:2025tgm, DeLuca:2025nao}.

PBHs may form through a variety of mechanisms, including the collapse of primordial adiabatic~\cite{Niemeyer:1999ak, Shibata:1999zs, Musco:2004ak, Uehara:2024yyp, Joana:2025gqf} and isocurvature~\cite{Passaglia:2021jla, Yoo:2021fxs} perturbations, (p)reheating dynamics~\cite{Green:2000he, Jedamzik:2010dq, Cotner:2018vug, Martin:2019nuw, Kou:2019bbc, Martin:2020fgl, Aurrekoetxea:2023jwd, Padilla:2025bkv, Kasai:2025coe, Adshead:2025gka}, scalar condensate fragmentation~\cite{Cotner:2016cvr, Cotner:2017tir, Cotner:2019ykd, Flores:2021jas}, primordial magnetic field~\cite{Maiti:2025ijr} or magneto-hydrodynamics~\cite{Liang:2025pll}, and collapse of topological defects such as cosmic strings~\cite{Hawking:1987bn, Polnarev:1988dh, Garriga:1992nm, Caldwell:1995fu, MacGibbon:1997pu, Helfer:2018qgv, Vilenkin:2018zol, Jenkins:2020ctp} and domain walls~\cite{Widrow:1989vj, Rubin:2000dq, Rubin:2001yw, Tanahashi:2014sma, Deng:2016vzb, Ge:2019ihf, Liu:2019lul, Kitajima:2025shn}, and cosmological first-order phase transitions (FOPTs). PBH formation associated with FOPTs can proceed through several distinct channels depending on the nature of the transition and the dynamics of vacuum bubbles (See Ref.~\cite{Banerjee:2025qji} for a mini-review): bubble collisions~\cite{Hawking:1982ga, Crawford:1982yz, Moss:1994iq, Khlopov:1998nm, Khlopov:1999ys, Freivogel:2007fx, Lewicki:2019gmv, Jung:2021mku}, collapse of bubbles nucleated during inflation~\cite{Garriga:2015fdk, Deng:2017uwc, Kusenko:2020pcg, Maeso:2021xvl, Escriva:2023uko, Huang:2023chx, Huang:2023mwy, Wang:2025hwc}, particles trapped in the false vacuum~\cite{Baker:2021nyl, Baker:2021sno, Lewicki:2023mik} (including collapsing Fermi balls~\cite{Kawana:2021tde, Gross:2021qgx, Kawana:2022lba, Huang:2022him, Marfatia:2022jiz} and Q-balls~\cite{Dent:2025lwe, Li:2026yhv}), interrupted transitions during an early matter-dominated reheating stage~\cite{Ai:2024cka}, inverted bubble collapse~\cite{Murai:2025hse}, and collapse of delayed-decayed false vacuum patches~\cite{Liu:2021svg, Hashino:2021qoq, He:2022amv, Hashino:2022tcs, Kawana:2022olo, Lewicki:2023ioy, Gouttenoire:2023naa, Baldes:2023rqv, He:2023ado, Jinno:2023vnr, Gouttenoire:2023pxh, Conaci:2024tlc, Lewicki:2024ghw, Kanemura:2024pae, Cai:2024nln, Banerjee:2024cwv, Lewicki:2024sfw, Wu:2024lrp, Ghoshal:2025dmi, Zou:2025sow, Kierkla:2025vwp, Zhang:2025kbu, An:2026hiq}.

The last mechanism~\cite{Liu:2021svg}---PBH formations via collapse of delayed-decayed false vacuum patches---arises from the stochastic and asynchronous nature of bubble nucleation during FOPTs. In particular, if the transition is strongly supercooled, bubble nucleation may be temporarily delayed in large regions, producing isolated symmetry-unbroken regions, which we refer to as the trapped false vacuum~\cite{Kodama:1981gu} or simply the false vacuum domains (FVDs)~\cite{Lewicki:2023ioy, Hashino:2025fse, Cao:2025jwb}. We refer to FOPTs that leave such large FVDs for extended periods as \emph{delayed} FOPTs. Outside an FVD, the universe is in the true vacuum and radiation-dominated; inside an FVD, the energy density contains both radiation and false vacuum energy components. As the universe expands, the radiation is diluted rapidly while the false vacuum energy inside FVD remains constant and larger than the true vacuum energy outside FVD, so a local overdensity develops inside the FVD that can potentially collapse to a PBH.

To determine whether a PBH could form from a delayed-decayed region, a commonly used criterion is a threshold for the local density contrast~\cite{Liu:2021svg, Hashino:2021qoq, He:2022amv, Hashino:2022tcs, Kawana:2022olo, Gouttenoire:2023naa, Baldes:2023rqv, He:2023ado, Gouttenoire:2023pxh, Conaci:2024tlc, Lewicki:2024ghw, Kanemura:2024pae, Cai:2024nln, Banerjee:2024cwv, Lewicki:2024sfw, Wu:2024lrp, Ghoshal:2025dmi, Zou:2025sow, Kierkla:2025vwp, Zhang:2025kbu}: if the density contrast exceeds a critical value $\delta_c$, the FVD is expected to collapse into a PBH. A frequently quoted value is $\delta_c \sim 0.45$, which originates from numerical simulations of PBH formation from curvature perturbations reentering the horizon during radiation domination~\cite{Niemeyer:1999ak, Shibata:1999zs, Musco:2004ak, Harada:2013epa}. Whether the same condition $\delta > 0.45$ applies to PBH formation from delayed FOPTs remains debated. Ref.~\cite{Flores:2024lng} pointed out that the equations of motion for a Hubble-sized patch containing radiation but no curvature perturbation admit only expanding solutions, and argued that a Schwarzschild criterion~\footnote{The Schwarzschild criterion can be generalized to the hoop conjecture~\cite{Klauder:1972je, Misner:1973prb, Ye:2025wif}, which accommodates non-spherical collapse. See also Ref.~\cite{Saini:2017tsz} for a modified hoop conjecture in an expanding spacetime.}---requiring the object to be localized within its Schwarzschild radius---should be used instead~\cite{Baker:2021sno, Lewicki:2023ioy, Jinno:2023vnr, An:2026hiq}. In addition, Refs.~\cite{Franciolini:2025ztf,Wang:2026zvz} suggested that the gauge dependence of the density contrast may substantially reduce the predicted PBH abundance for delayed FOPTs. To clarify these issues, we perform fully nonlinear simulations of a coupled gravity-scalar-fluid system to study the collapse dynamics of FVDs and to examine PBH formation criteria in detail.

A similar process of PBH formation occurs in the collapse of bubbles nucleated during inflation. These true-vacuum bubbles during inflation become false-vacuum bubbles after inflation ends, and their final fate depends on the bubble mass~\cite{Garriga:2015fdk, Deng:2017uwc}. Bubbles below a critical mass collapse to a singularity and form a normal (``subcritical'') black hole, while heavier bubbles inflate internally to produce a baby universe connected to the exterior FLRW region by a wormhole that eventually pinches off. Observers outside the baby universe then see a PBH, which is referred to as a ``supercritical'' black hole. For PBHs formed from primordial curvature perturbations, a recent classification distinguishes between the type A and type B PBHs according to the absence or presence of bifurcating trapping horizons (where both outgoing and ingoing null expansions vanish, $\Theta_+ = \Theta_- = 0$)~\cite{Uehara:2024yyp}~\footnote{Some literatures also refer to type A/B PBHs as type I/II PBHs. Actually, type I/II is a classification of curvature perturbations based on whether the areal radius is a monotonic function of the radial coordinate, and type II perturbations do not necessarily form type B PBHs in the radiation-dominated background.}. The intersection of these expansions locates a wormhole throat at which the physical radius attains a local minimum, so this classification resembles the subcritical/supercritical distinction. As shown in Ref.~\cite{Deng:2016vzb}, a bifurcating trapping horizon does form during supercritical PBH formation. In this paper, we adopt the type A/B classification since it is straightforward to distinguish them numerically by checking for bifurcating trapping horizons.

The investigation of FVD dynamics has a long history. Most studies use Israel's junction conditions to match interior and exterior metrics and to follow the motion of the FVD boundary, under the assumptions of spherical symmetry and negligible wall thickness. In early works~\cite{Kodama:1981gu, Sato:1981bf, Sato:1981gv, Kodama:1982sf, Berezin:1982ur, Blau:1986cw}, an interior de Sitter region was matched to an exterior Schwarzschild metric without including the radiation fluid. More recent analyses~\cite{Hashino:2025fse, Cao:2025jwb, Dent:2025bwo} have incorporated the radiation fluid, which is more realistic for FOPTs in the early Universe. Two kinds of solutions can be obtained from the equation of motion of the FVD boundary. If the FVD boundary radius tends to infinity, a baby universe is realized, and a supercritical PBH is formed. Ref.~\cite{Hashino:2025fse} used this approach and found that the ratio between the characteristic timescales for horizon crossing and vacuum-energy domination can serve as a criterion for supercritical PBH formation. However, the thin-wall approximation cannot distinguish whether a contracting FVD ultimately produces a subcritical PBH or simply disperses as its radius reaches zero. Moreover, computing PBH abundances from delayed FOPTs requires accounting for subcritical PBHs, which can form via direct collapse of shrinking bubble walls~\cite{Flores:2024lng} or through subsequent accretion of ambient radiation~\cite{Tanahashi:2014sma}. These dynamics cannot be captured by Israel's junction conditions, \textit{motivating} fully nonlinear simulations to resolve the detailed dynamics of FVD collapse and to establish robust PBH formation criteria. For simplicity, we also assume spherical symmetry and that no bubble nucleation occurs inside the FVD, which is a reasonable approximation for strongly supercooled transitions. Our simulation setup also accommodates other FVD collapse scenarios, such as the collapse of inflationary bubbles~\cite{Garriga:2015fdk, Deng:2017uwc, Maeso:2021xvl, Wang:2025hwc} and inverted bubbles~\cite{Murai:2025hse}, which naturally admit spherical symmetry.

However, there are important differences between our setup and previous simulations of inflationary bubble collapse in Ref.~\cite{Deng:2017uwc}. In that paper, the bubble field was assumed to be strongly coupled to Standard Model particles so that the bubble wall is effectively impermeable~\footnote{See Ref.~\cite{Deng:2016vzb} for simulations of permeable domain walls.}, the fluid is confined to the exterior, and the bubble interior is described by de Sitter spacetime. For permeable walls, only analytical estimates of the PBH mass exist in certain limits~\cite{Deng:2020mds}. In the present work, we assume permeable walls that interact with the radiation fluid only through gravity, and thus our FVDs are filled with radiation.

This paper is organized as follows. In Sec.~\ref{sec:setup}, we introduce the simulation setup, including the model and equations of motion, initial and boundary conditions, and numerical methods. In Sec.~\ref{sec:dynamics}, we present the collapse dynamics of FVDs and classify outcomes as type B (supercritical), type A (subcritical), and no PBH formation. In Sec.~\ref{sec:thresholds}, we examine several PBH formation criteria for FVD collapse and compare them with our simulation results. Finally, we conclude in Sec.~\ref{sec:conclusions}. Throughout this paper, we adopt geometrical units with $c = G = 1$.

\section{Numerical simulation setups} \label{sec:setup}

\subsection{The model and equations of motion}

We consider a coupled gravity-scalar-fluid system with the action
\begin{equation}
    S = \int \mathrm{d}^4x \sqrt{-g} \left[\frac{{}^{(4)}R}{16\pi} - \frac{1}{2} g^{\mu\nu} \partial_\mu \phi \partial_\nu \phi - V(\phi) + \mathcal{L}_\mathrm{fluid}\right],
\end{equation}
where ${}^{(4)}R$ is the four-dimensional Ricci scalar, $g_{\mu\nu}$ is the metric tensor, $\phi$ is a scalar field with a potential $V(\phi)$, and $\mathcal{L}_\mathrm{fluid}$ is the Lagrangian density of a perfect fluid. As stated above, we assume spherical symmetry for the FVD so that the spacetime metric can be written as
\begin{equation} \label{eq:metric}
    \mathrm{d}s^2=-A(t,r)^2\mathrm{d}t^2+B(t,r)^2\mathrm{d}r^2+R(t,r)^2\mathrm{d}\Omega^2,
\end{equation}
where $A$ is the lapse function, $B$ is the radial metric component, $R$ is the areal radius, and $\mathrm{d}\Omega^2=\mathrm{d}\theta^2+\sin^2\theta \mathrm{d}\varphi^2$ is the two-sphere line element. Following the notation of Ref.~\cite{Misner:1964je}, we define the auxiliary quantities
\begin{subequations}
  \begin{align}
    D_tR &\equiv \frac{1}{A}\frac{\partial R}{\partial t} \equiv U(t,r),\\
    D_rR &\equiv \frac{1}{B}\frac{\partial R}{\partial r} \equiv \Gamma(t,r),
  \end{align}
\end{subequations}
where $D_t$ and $D_r$ are the proper time and distance derivatives, respectively. Here $U$ represents the radial component of the fluid four-velocity in the associated Eulerian frame, and $\Gamma$ is a generalization of the Lorentz factor. To handle possible non-monotonic behavior of $R$ during evolution, we also introduce~\cite{Deng:2016vzb, Escriva:2025eqc}
\begin{equation}
    K \equiv -\frac{1}{A}\left(\frac{\dot{B}}{B}+\frac{2\dot{R}}{R}\right),
\end{equation}
which is the trace of the extrinsic curvature, $K \equiv \gamma^{ij}K{ij}$, of the spatial metric $\gamma_{ij}$ associated with Eq.~\eqref{eq:metric}. Here dots denote partial derivatives with respect to $t$ and $K_{ij} \equiv -\mathcal{L}_n\gamma_{ij}/2$ with $\mathcal{L}_n$ the Lie derivative along the unit normal vector $n^\mu = (1/A, 0, 0, 0)$ orthogonal to constant-$t$ hypersurfaces. In a spatially flat Friedmann-Lema\^{i}tre-Robertson-Walker (FLRW) background, $K$ reduces to $K_b = -3H$, so $K$ encodes the local Hubble expansion, including the modification from cosmological fluctuations. The three-dimensional Ricci scalar of the spatial slices reads
\begin{equation}
    {}^{(3)}R = \frac{2}{R^2}\left[1 - \frac{(R')^2}{B^2} - \frac{2R}{B}\left(\frac{R'}{B}\right)'\right],
\end{equation}
where primes denote partial derivatives with respect to $r$. The detailed derivations of the extrinsic curvature and ${}^{(3)}R$ are given in Appendix~\ref{app:K}.

The energy-momentum tensors of the scalar field and the fluid are
\begin{subequations}
    \begin{align}
        T_{\mu\nu}^{(\phi)} &= \partial_\mu \phi \partial_\nu \phi - g_{\mu\nu} \left( \frac{1}{2} g^{\alpha\beta} \partial_\alpha \phi \partial_\beta \phi + V(\phi) \right), \\
        T_{\mu\nu}^{(\mathrm{fluid})} &= (\rho + p) u_\mu u_\nu + p g_{\mu\nu},
    \end{align}
\end{subequations}
where $\rho$, $p$, and $u^\mu$ are the fluid energy density, pressure, and four-velocity, respectively. The total energy-momentum tensor is $T_{\mu\nu} = T_{\mu\nu}^{(\phi)} + T_{\mu\nu}^{(\mathrm{fluid})}$. We assume only gravitational interaction between the scalar field and the fluid, so that each energy-momentum tensor is separately conserved,
\begin{subequations}
    \begin{align}
        \nabla^\mu T_{\mu\nu}^{(\phi)} &= 0, \\
        \nabla^\mu T_{\mu\nu}^{(\mathrm{fluid})} &= 0.
    \end{align}
\end{subequations}
The fluid equation of state is taken to be $p = \omega\rho$ with a constant $\omega$. Since we focus on FOPTs during the radiation-dominated era, we set $\omega = 1/3$. The fluid four-velocity can be parameterized as
\begin{equation}
    u^{\mu} = \left(\frac{1}{A\sqrt{1-v^{2}}},\frac{v}{B\sqrt{1-v^{2}}},0,0\right)
\end{equation}
where $v(t,r)$ is the fluid three-velocity relative to the comoving radial coordinate $r$.

There is gauge freedom in the choice of the lapse function $A(t,r)$. In the conventional Misner-Sharp formalism~\cite{Misner:1964je}, the lapse function is determined by the condition that the fluid three-velocity vanishes, $v(t,r) = 0$, which is called comoving slicing. However, in our scenario, we find it more numerically stable to choose the gauge condition $A(t,r) = 1$, which is called geodesic slicing and implies that the coordinate time $t$ corresponds to the proper time of observers at fixed spatial coordinates. In this gauge, the fluid three-velocity $v(t,r)$ is nonzero in general. In what follows, we therefore set $A(t,r) = 1$.

\begin{figure}[htbp]
    \centering
    \includegraphics[width=0.6\linewidth]{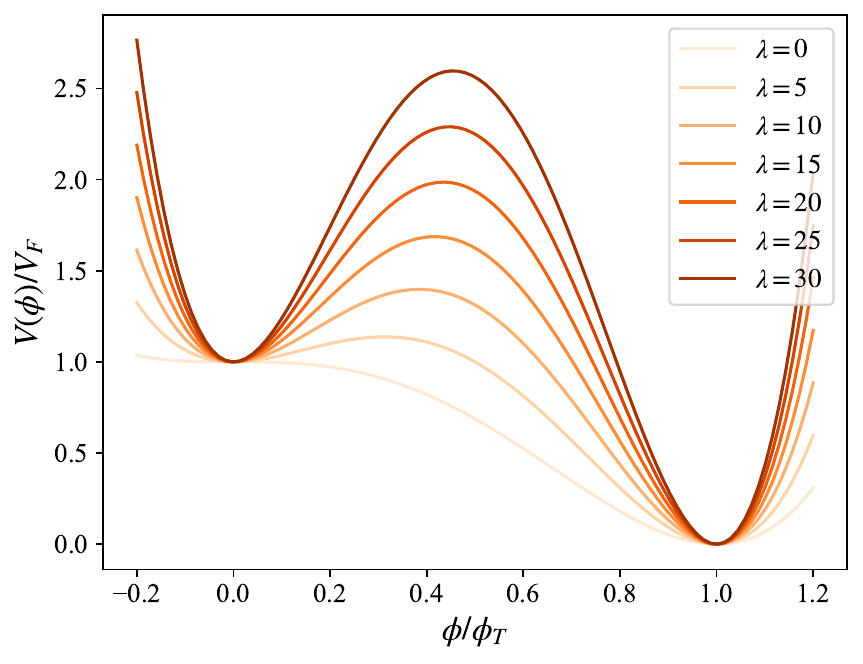}
    \caption{The scalar potential $V(\phi)$ for different values of $\lambda$. The potential is normalized by $V_\mathrm{F}$ and the field by $\phi_\mathrm{T}$.}
    \label{fig:potential}
\end{figure}

To realize a FOPT, we adopt a temperature-independent scalar potential for simplicity~\cite{Giblin:2013kea, Giblin:2014qia, Lewicki:2019gmv,Yuwen:2024gcf} as follows,
\begin{equation}
    V(\phi) = \left[1 + \lambda \left(\frac{\phi}{\phi_\mathrm{T}}\right)^2 - (2\lambda + 4)\left(\frac{\phi}{\phi_\mathrm{T}}\right)^3 + (\lambda + 3)\left(\frac{\phi}{\phi_\mathrm{T}}\right)^4\right] \left(V_\mathrm{F} - V_\mathrm{T}\right) + V_\mathrm{T},
\end{equation}
where $\phi_\mathrm{T}$ denotes the true-vacuum field value (the false vacuum is at $\phi = 0$), $V_\mathrm{F}$ and $V_\mathrm{T}$ are the potential energy densities at the false and true vacua, respectively, and $\lambda$ is a dimensionless parameter that controls the position and height of the barrier between the two vacua, as illustrated in Fig.~\ref{fig:potential}. In the limit of $\lambda \rightarrow +\infty$, the minima of the potential become nearly degenerate compared to the barrier, and this corresponds to the thin-wall limit. In the limit of $\lambda \rightarrow 0$, the potential barrier becomes infinitesimally small relative to the potential energy difference, which is called the thick-wall limit. The energy difference between the false and true vacua is
\begin{equation}
    \rho_\mathrm{V} = V_\mathrm{F} - V_\mathrm{T},
\end{equation}
and in this work we set $V_\mathrm{T} = 0$, so that $\rho_\mathrm{V} = V_\mathrm{F}$.

In the thin-wall limit, the bubble wall thickness is estimated by~\cite{Cutting:2018tjt, Cutting:2020nla, Li:2025zxa}
\begin{equation}
    l^{\mathrm{tw}} = \frac{2}{\sqrt{\partial_{\phi}^2V(\phi_\mathrm{T})}} = \frac{\sqrt{2}\phi_\mathrm{T}}{\sqrt{(\lambda+6)\rho_\mathrm{V}}},
\end{equation}
and the surface tension can be approximated as~\cite{Maeso:2021xvl, Wang:2025eee, Cao:2025jwb}
\begin{equation} \label{eq:sigma}
    \sigma \simeq \int_0^{\infty}\mathrm{d}r\left[\frac{1}{2}\left(\frac{\mathrm{d}\phi}{\mathrm{d}r}\right)^2 + V(\phi) - V_\mathrm{T}\right] \simeq \int_0^{\phi_\mathrm{T}} \mathrm{d}\phi \sqrt{2(V(\phi) - V_\mathrm{T})} \simeq \frac{\lambda^\frac{3}{2}\phi_\mathrm{T}\sqrt{2\rho_\mathrm{V}}}{6(\lambda+3)}.
\end{equation}

The scalar-field energy density $\rho_\phi = \rho_\mathrm{K} + \rho_\mathrm{V} + \rho_\mathrm{G}$ decomposes into kinetic, gradient, and potential parts,
\begin{align}
    \rho_\mathrm{K} &= \frac{\dot{\phi}^2}{2A^2},\\
    \rho_\mathrm{G} &= \frac{\phi'^2}{2B^2},\\
    \rho_\mathrm{V} &= V(\phi).
\end{align}
The total energy density of the system is
\begin{equation}
    \rho_\mathrm{tot} = \rho + \rho_\phi = \rho + \rho_\mathrm{K} + \rho_\mathrm{G} + \rho_\mathrm{V}.
\end{equation}

For numerical convenience, we introduce dimensionless variables by rescaling with the initial Hubble parameter $H_i$,
\begin{equation}
    \tilde{t} = H_i t, \quad \tilde{r} = H_i r, \quad \tilde{R} = H_i R, \quad \tilde{K} = \frac{1}{H_i} K, \quad \tilde{\rho} = \frac{1}{H_i^2} \rho, \quad \tilde{V} = \frac{1}{H_i^2} V,
\end{equation}
and thereafter drop tildes for notational simplicity. With this choice, the initial Hubble parameter is $H_i = 1$, the initial time is $t_i = \alpha$ with $\alpha = 2/(3(1+\omega))$, and the initial background energy density is $\rho_b(t_i) = 3/(8\pi)$.

In these dimensionless variables, the Einstein equations together with the conservation equations yield
\begin{subequations}
    \begin{align}
        \label{eq:R} \dot{R} &= U, \\
        \label{eq:U} \dot{U} &= -\frac{1+U^{2}-\Gamma^{2}}{2R}-4\pi R{T^{1}}_{1}, \\
        \label{eq:rho} \dot{\rho} &= \frac{(1+\omega)\rho}{1-\omega v^{2}}\left[-v^{2}\left(K+\frac{2U}{R}\right)+K-\frac{2v\Gamma}{R}-\frac{v^{\prime}}{B}\right]-\frac{1-\omega}{1-\omega v^{2}}\frac{\rho^{\prime}v}{B}, \\
        \label{eq:v} \dot{v} &= \frac{(1-v^{2})v}{1-\omega v^{2}}\left[\left(K+\frac{2U}{R}\right)-\omega K+\frac{2\omega v\Gamma}{R}-\frac{(1-v^{2})\omega}{(1+\omega)v}\frac{\rho^{\prime}}{\rho B}\right]-\frac{1-\omega}{1-\omega v^{2}}\frac{v^{\prime}v}{B}, \\
        \label{eq:B} \dot{B} &= -B\left(K+\frac{2U}{R}\right), \\
        \label{eq:Gamma} \dot{\Gamma} &= -\frac{4\pi RT_{01}}{B}, \\
        \label{eq:K} \dot{K} &= \left(K+\frac{2U}{R}\right)^{2}+2\left(\frac{U}{R}\right)^{2}+4\pi(T_{00}+{T^{1}}_{1}+2{T^{2}}_{2}), \\
        \label{eq:phi} \dot{\phi} &= \Pi, \\
        \label{eq:Pi} \dot{\Pi} &= K\Pi+\frac{1}{BR^{2}}\left(\frac{R^{2}}{B}\phi^{\prime}\right)^{\prime}-\partial_{\phi}V,
    \end{align}
\end{subequations}
where $\Pi \equiv \dot{\phi}$ and the stress-tensor components are
\begin{subequations}
    \begin{align}
        T_{00} &= \frac{1+\omega v^{2}}{1-v^{2}}\rho+\frac{1}{2}\dot{\phi}^{2}+\frac{1}{2B^2}\phi^{\prime2}+V(\phi), \\
        {T^{1}}_{1} &= \frac{\omega+v^{2}}{1-v^{2}}\rho+\frac{1}{2}\dot{\phi}^{2}+\frac{1}{2B^2}\phi^{\prime2}-V(\phi), \\
        {T^{2}}_{2} &= \omega\rho+\frac{1}{2}\dot{\phi}^{2}-\frac{1}{2B^2}\phi^{\prime2}-V(\phi), \\
        T_{01} &= \dot{\phi}\phi^{\prime}-\frac{1+\omega}{1-v^{2}}\rho vB.
    \end{align}
\end{subequations}
These equations form the basis of our numerical evolution described below.

\subsection{Initial conditions}

We specify initial data at $t = t_i$ as follows. The initial scale factor is set to $a_i = 1$. We model the system as an FVD embedded in a radiation-dominated FLRW background and therefore initialize metric components and auxiliary variables to their FLRW values as
\begin{equation}
    B(t_i,r) = 1, \quad R(t_i,r) = r, \quad U(t_i,r) = r, \quad \Gamma(t_i,r) = 1, \quad K(t_i,r) = -3.
\end{equation}

The scalar field transitions smoothly from the false vacuum $\phi = 0$ at the FVD center to the true vacuum $\phi = \phi_\mathrm{T}$ at large $r$ with an illustrative profile of the form
\begin{equation}
    \phi(t_i,r) = \frac{\phi_\mathrm{T}}{2}\left[1+\tanh\left(\frac{r-r_i}{l}\right)\right],
\end{equation}
where $r_i$ is the initial radius of the FVD and $l$ is the initial wall thickness. In this work, we take
\begin{equation}
    l = \frac{r_i}{10}.
\end{equation}
We have verified that moderate variations of $l$ do not qualitatively affect the subsequent evolution. The scalar field is taken to be initially at rest, $\Pi(t_i, r) = 0$, in accordance with Ref.~\cite{Hashino:2025fse}. 

The radiation fluid is initially comoving, $v(t_i,r) = 0$. To satisfy the Hamiltonian constraint on the initial time slice, we choose the initial fluid energy density to compensate the scalar-field contribution~\cite{Deng:2016vzb},
\begin{equation}
    \rho(t_i,r) = \rho_b(t_i) - \rho_\phi(t_i,r) = \frac{3}{8\pi} - \frac{1}{2}\phi'^2(t_i,r) - V(\phi(t_i,r)),
\end{equation}
where $\rho_b(t)$ denotes the FLRW background radiation energy density at time $t$. \black{By setting the initial fluid energy density in this way, the total energy density is homogeneous, $\rho_\mathrm{tot}(t_i,r) = \rho_b(t_i)$, and the curvature perturbation is zero on our chosen initial hypersurface. This implies that the initial configuration in our simulations is a compensated isocurvature perturbation between the scalar/vacuum sector and the radiation fluid. The transition from this isocurvature perturbation to a density one is dynamically realized during the evolution: as the universe expands, the exterior radiation dilutes while the interior vacuum energy remains constant, actively generating a local total density contrast.}

\subsection{Boundary conditions}

At the origin $r = 0$, obviously we have $R(t,0) = 0$ and $U(t,0) = 0$. Regularity imposes $\phi'(t,0) = 0$, $\Pi'(t,0) = 0$, $\rho'(t,0) = 0$, and $v(t,0) = 0$. By parity symmetry, the metric functions $A, B, \Gamma, K$ are even functions of $r$.

At the outer boundary $r = r_\mathrm{max}$, the scalar field is fixed in the true vacuum: $\phi(t,r_\mathrm{max}) = \phi_\mathrm{T}$ and $\Pi(t,r_\mathrm{max}) = 0$, so that the spacetime approaches a pure-radiation FLRW solution. In our simulations, we set $r_\mathrm{max} = 5 r_i$, which places the outer boundary sufficiently far from the FVD so that it remains effectively in the FLRW regime throughout the evolution.

\subsection{Misner-Sharp mass and expansions}

We characterize the gravitational energy contained within a sphere of areal radius $R$ using the Misner-Sharp mass~\cite{Misner:1964je, May:1966zz, Hayward:1994bu}, defined for a spherically symmetric metric by
\begin{equation}
    M = \frac{R}{2}\left(1-g^{\mu\nu}\partial_{\mu}R\partial_{\nu}R\right).
\end{equation}
In our metric ansatz, this becomes
\begin{equation} \label{eq:MSmass}
    M = \frac{R}{2}(1+U^{2}-\Gamma^{2}).
\end{equation}
From the Einstein equations, one obtains
\begin{equation} \label{eq:error}
    M' = 4\pi R^{2}\left(R'T_{00}-\dot{R}T_{01}\right),
\end{equation}
which is actually a combination of the Hamiltonian and momentum constraints and provides a useful monitor of numerical accuracy (see Appendix~\ref{app:convergence}).

The local causal geometry of a spherical surface $r = \mathrm{const}$ can be characterized by the expansions of the outgoing and ingoing null geodesic congruences orthogonal to the surface. After introducing the outgoing ($+$) and ingoing ($-$) null vectors,
\begin{equation}
  k_\mu^\pm = \frac{1}{\sqrt{2}}(-A,\pm B,0,0),
\end{equation}
satisfying $k^+ \cdot k^- = -1$, the expansions are defined by
\begin{equation}
  \Theta^\pm \equiv h^{\mu\nu}\nabla_\mu k_\nu^\pm,
\end{equation}
where $h^{\mu\nu}$ is the induced metric on the two-sphere. In our coordinates, this reduces to
\begin{equation} \label{eq:expansion}
  \Theta^\pm = \frac{\sqrt{2}}{R}(U\pm\Gamma).
\end{equation}

We classify surfaces according to the signs of $\Theta^\pm$: a \emph{normal} surface has $\Theta^+ > 0$ and $\Theta^- < 0$; a \emph{trapped} surface has $\Theta^\pm < 0$; and an \emph{anti-trapped} surface has $\Theta^\pm > 0$. The transition from a normal to a trapped surface should satisfy $\Theta^+ = 0$ and $\Theta^- < 0$, which corresponds to a marginally trapped surface, usually called an \emph{apparent horizon}. Conversely, the transition from a normal to an anti-trapped surface should satisfy $\Theta^- = 0$ and $\Theta^+ > 0$, which corresponds to a marginally anti-trapped surface, and possibly being a cosmological or white-hole horizon. A \emph{bifurcating trapping horizon} occurs when $\Theta^+ = \Theta^- = 0$ simultaneously. We adopt the formation of an apparent horizon or a bifurcating trapping horizon as the diagnostic for type A and type B PBH formation, respectively. 

\black{As we will see in our simulations, PBH formation is not only imposed or checked at the horizon-crossing time $t_\mathrm{H}$. Instead, we continuously evolve the fully nonlinear gravity-scalar-fluid system from an initially super-horizon FVD, through horizon crossing, and into the sub-horizon regime, until either a trapping horizon forms or the FVD disperses. It is the actual appearance of an apparent horizon or a bifurcating trapping horizon that diagnoses PBH formation. In spherical symmetry, this horizon-based diagnostic is equivalent to the Schwarzschild condition $2M/R = 1$. Numerically, we continuously check whether a black hole horizon forms during the full evolution process. Therefore, our simulations are capable of capturing PBHs that form before, around, or after the nominal horizon-crossing time.}

\subsection{Numerical implementation}

We evolve Eqs.~\eqref{eq:R}--\eqref{eq:Pi} using the method of lines. Radial derivatives are discretized using fourth-order finite-difference stencils on a uniform grid of $N$ points, and time integration is performed with a classical fourth-order Runge-Kutta scheme. Ghost zones are added at both boundaries to impose boundary conditions. The time step $\Delta t$ is chosen to satisfy the Courant-Friedrichs-Lewy (CFL) condition, $\Delta t = C \Delta r$ with $C < 1$.

To resolve the shrinking comoving thickness of bubble walls in an expanding spacetime, we implement a Berger-Oliger adaptive mesh refinement (AMR) scheme~\cite{Berger:1984zza} with the scalar-field gradient $|\phi'|$ as the refinement indicator. Interpolation between levels (radial and temporal) is performed linearly; time stepping on each level is adjusted to satisfy the CFL condition for that level. Recent applications of similar AMR strategies in numerical relativity and cosmology can be found in Refs.~\cite{Wainwright:2013lea, Yoo:2024lhp, Kitajima:2025shn}.

To suppress high-frequency numerical errors introduced by finite-difference truncation and regridding, we add a Kreiss-Oliger dissipation term~\cite{1972Tell...24..199K} to the evolution of every dynamical variable $f$:
\begin{equation}
    \partial_t f_m \rightarrow \partial_t f_m + \frac{\epsilon}{64\Delta r}(f_{m+3} - 6f_{m+2} + 15f_{m+1} - 20f_m +15f_{m-1} - 6f_{m-2} + f_{m-3}),
\end{equation}
where $m \pm n$ labels the grid point offset and $\epsilon$ is an adjustable dissipation parameter typically of the order $\mathcal{O}(10^{-2})$.

Unless stated otherwise, our simulations use a base grid with $N = 5000$ points and a maximum refinement ratio of $16$. We take $C = 0.5$ for the CFL condition and $\epsilon = 0.01$ for the Kreiss-Oliger dissipation. Convergence tests of our numerical implementation are presented in Appendix~\ref{app:convergence}.

\section{Dynamics of false vacuum domains} \label{sec:dynamics}

The remaining dimensionless parameters in our simulations are $\phi_\mathrm{T}$, $\lambda$, $V_\mathrm{F}$, and $r_i$. For a representative parameter set, we choose
\begin{equation}
    \phi_\mathrm{T} = 0.1, \quad \lambda = 20, \quad r_i = 10,
\end{equation}
which corresponds to a relatively thin-wall regime. We then vary $V_\mathrm{F}$ to investigate its influence on the evolution of FVDs and on PBH formation. As discussed in Refs.~\cite{Garriga:2015fdk, Deng:2017uwc}, FVD collapse can produce two qualitatively different PBH outcomes depending on the competition between vacuum energy and gravitational collapse. If vacuum energy dominates the FVD dynamics, the interior inflates and a baby universe forms, yielding a supercritical (type B) PBH; if gravitational collapse dominates, the FVD directly collapses to a subcritical (type A) PBH without forming a baby universe. A third possibility is that the FVD disperses entirely and hence no PBH forms. Below, we present representative simulation results for these three outcomes.

\subsection{Type B PBH formation}

\begin{figure}[htbp]
    \centering
    \begin{subfigure}{0.495\linewidth}
        \includegraphics[width=\linewidth]{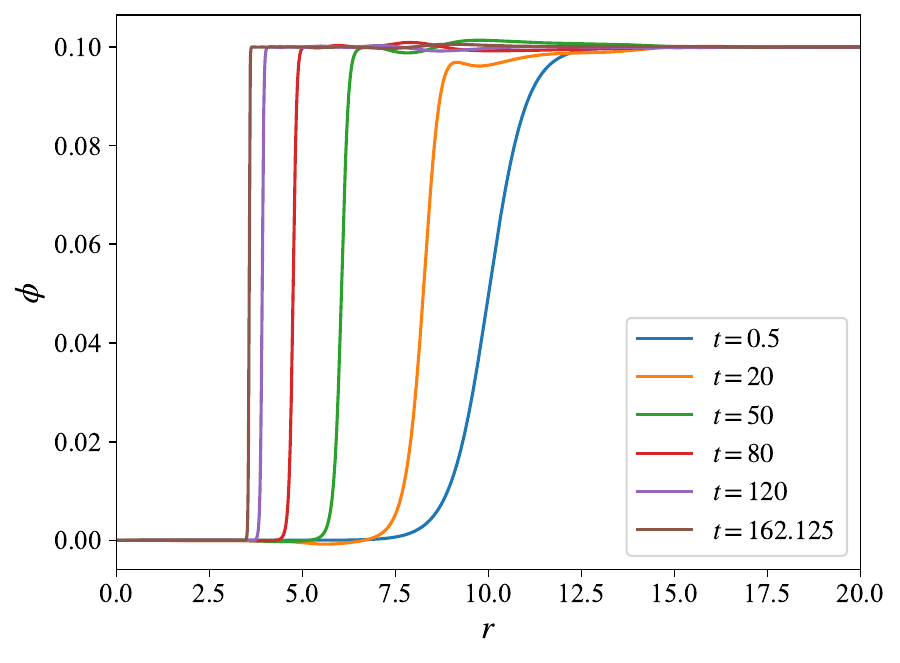}
        \caption{}
    \end{subfigure}
    \begin{subfigure}{0.495\linewidth}
        \includegraphics[width=\linewidth]{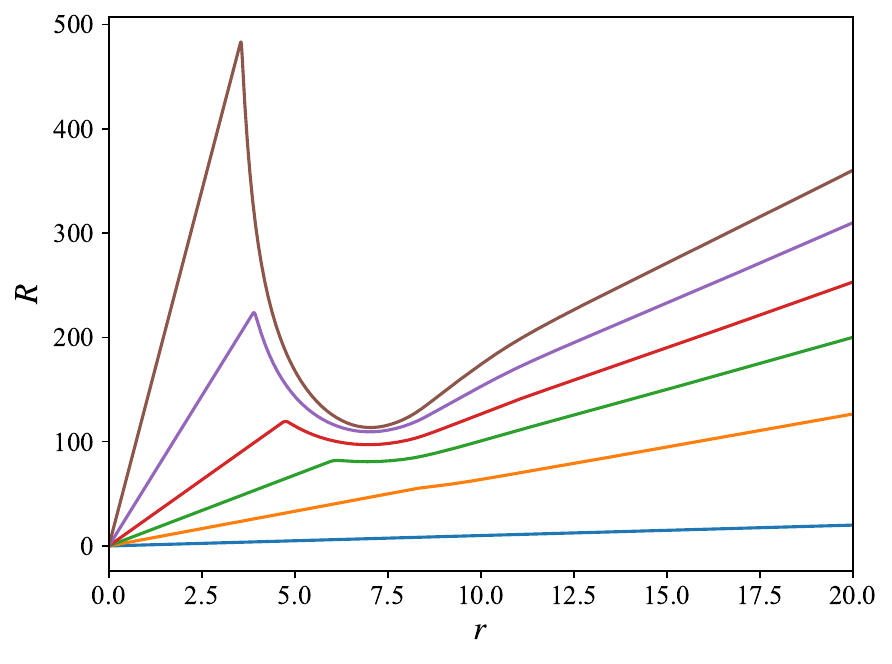}
        \caption{}
    \end{subfigure}
    \\
    \begin{subfigure}{0.495\linewidth}
        \includegraphics[width=\linewidth]{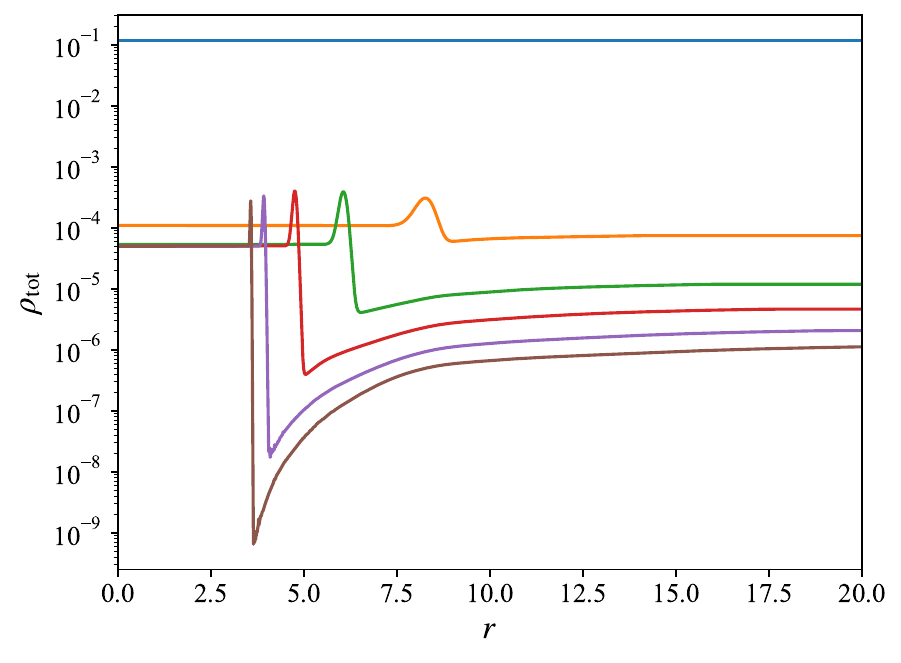}
        \caption{}
    \end{subfigure}
    \begin{subfigure}{0.495\linewidth}
        \includegraphics[width=\linewidth]{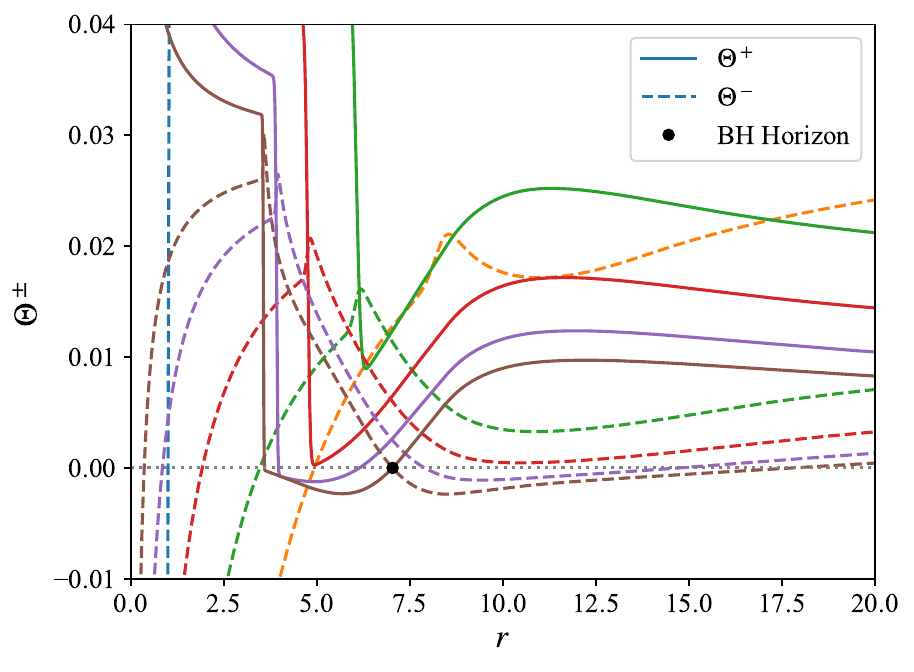}
        \caption{}
    \end{subfigure}
    \\
    \begin{subfigure}{0.495\linewidth}
        \includegraphics[width=\linewidth]{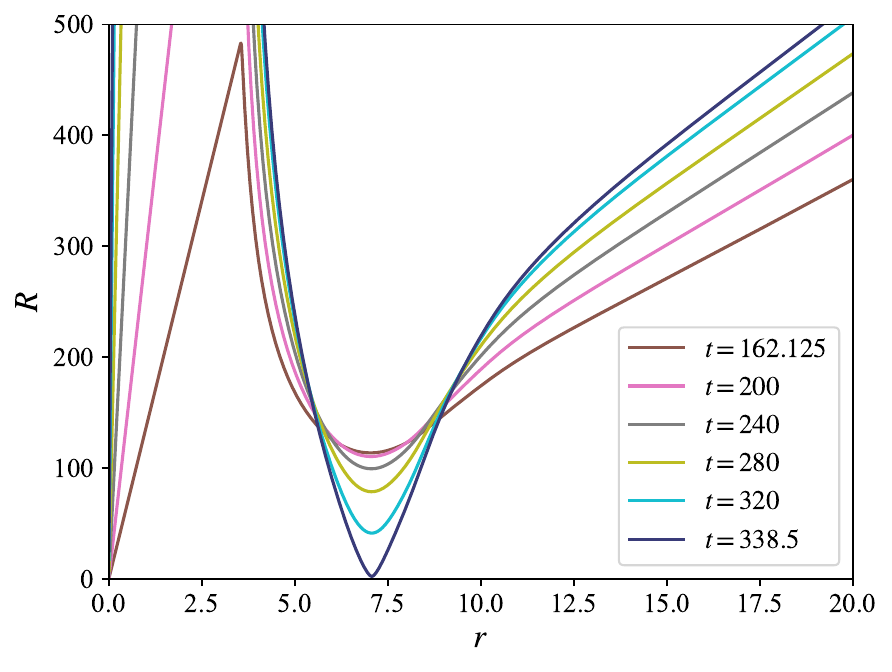}
        \caption{}
    \end{subfigure}
    \begin{subfigure}{0.495\linewidth}
        \includegraphics[width=\linewidth]{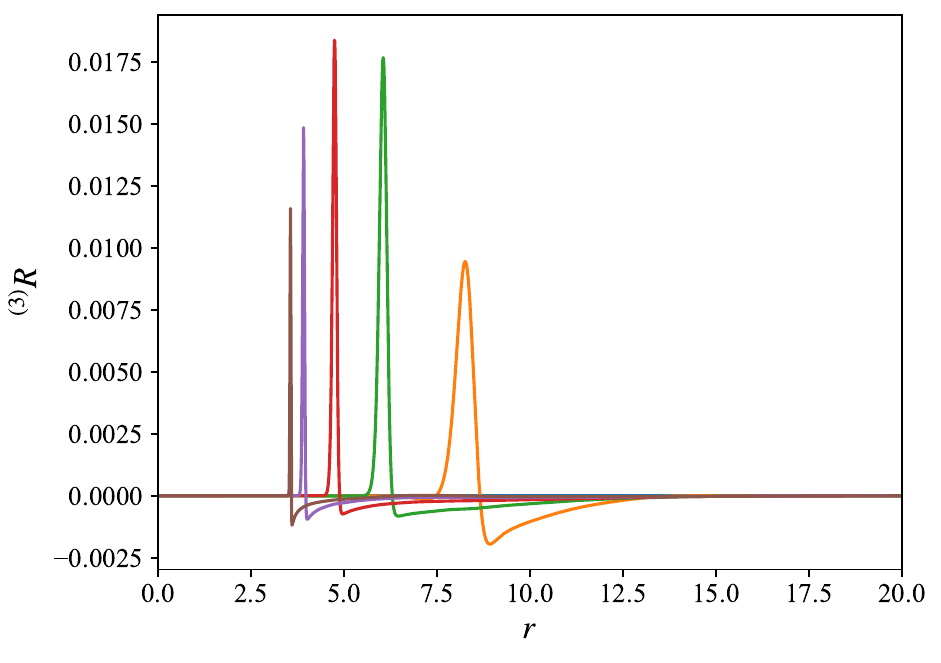}
        \caption{}
    \end{subfigure}
    \caption{Time evolution of several quantities for type B PBH formation with $V_\mathrm{F} = 5\times10^{-5}$. (a) Scalar-field profile $\phi$. (b) Areal radius $R$ before the formation of the bifurcating trapping horizon. (c) Total energy density $\rho_\mathrm{tot}$. (d) Expansions $\Theta^\pm$; the black dot marks the bifurcating trapping horizon. (e) Areal radius $R$ after horizon formation. (f) Three-dimensional Ricci scalar ${}^{(3)}R$.}
    \label{fig:typeB}
\end{figure}

We first consider type B PBH formation. Fig.~\ref{fig:typeB} shows the time evolution of several quantities for $V_\mathrm{F} = 5\times10^{-5}$ (the blue lines correspond to the initial time $t_i = \alpha = 1/2$). 

Panel (a) displays the scalar-field profile $\phi$ at different times: the FVD initially shrinks rapidly due to the pressure difference between the vacua, and when the FVD interior begins to inflate, the comoving radius of the wall decreases more slowly. Panel (b) shows the areal-radius profile $R$ before the formation of the bifurcating trapping horizon: the areal radius inside the FVD undergoes exponential expansion, and a local maximum develops at the bubble wall. A local minimum also appears outside the FVD, signaling the formation of a wormhole that connects the baby universe inside the FVD to the exterior universe. This throat coincides with the location where $\Theta^+$ and $\Theta^-$ intersect, and its areal radius increases during this stage.

Panel (c) presents the total energy density at different times. \black{At the initial time, the total energy density is homogeneous due to our choice of initial conditions. However, the local difference in the equation of state dynamically converts the initial isocurvature perturbation into a total density contrast.} Inside the FVD, the energy density rapidly approaches the false vacuum value $\rho_\mathrm{V} = V_\mathrm{F} = 5\times10^{-5}$ since the radiation fluid is diluted by the exponential expansion, while the vacuum energy remains constant. Outside the FVD, the radiation energy density decreases with cosmic expansion. This produces a large density contrast that increases over time, resulting in PBH formation. We also observe that the total energy density has a peak at the FVD boundary due to the bubble-wall energy.

At $t = 162.125$, a bifurcating trapping horizon forms at the wormhole throat, satisfying $\Theta^+ = \Theta^- = 0$; this is indicated by the black dot in panel (d). The bifurcating horizon then splits into two marginal surfaces: one with $\Theta^+ = 0$, $\Theta^- < 0$ (an apparent horizon) and the other with $\Theta^- = 0$, $\Theta^+ > 0$ (the corresponding anti-trapped branch); these two horizons then separate and move in opposite directions. After bifurcation, the areal radius of the throat begins to decrease and eventually reaches zero, indicating a singularity forms and the throat pinches off from the exterior universe (see panel (e)).

Panel (f) shows the three-dimensional Ricci scalar ${}^{(3)}R$. Initially, the Ricci scalar vanishes everywhere because the initial spacetime configuration is chosen to match the flat FLRW solution. As the bubble wall shrinks, a localized region of positive spatial curvature develops around the wall due to the concentrated energy. Therefore, even though the overdensity does not come from the inflationary curvature perturbation, the collapse of the FVD generates local curvature in the surrounding spacetime that provides a locally contracting solution and promotes PBH formation.

\subsection{Type A PBH formation}

\begin{figure}[htbp]
    \centering
    \begin{subfigure}{0.495\linewidth}
        \includegraphics[width=\linewidth]{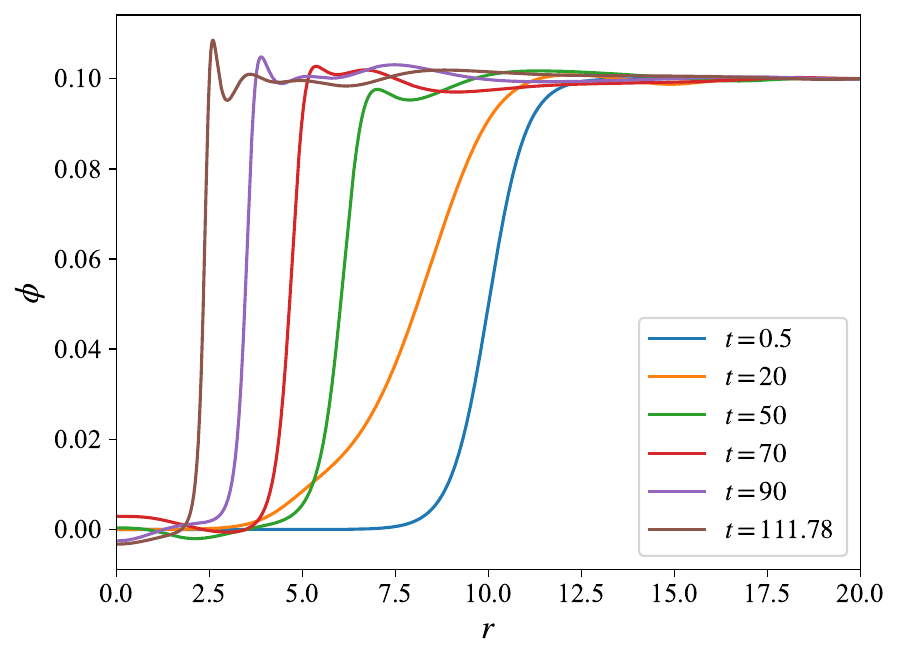}
        \caption{}
    \end{subfigure}
    \begin{subfigure}{0.495\linewidth}
        \includegraphics[width=\linewidth]{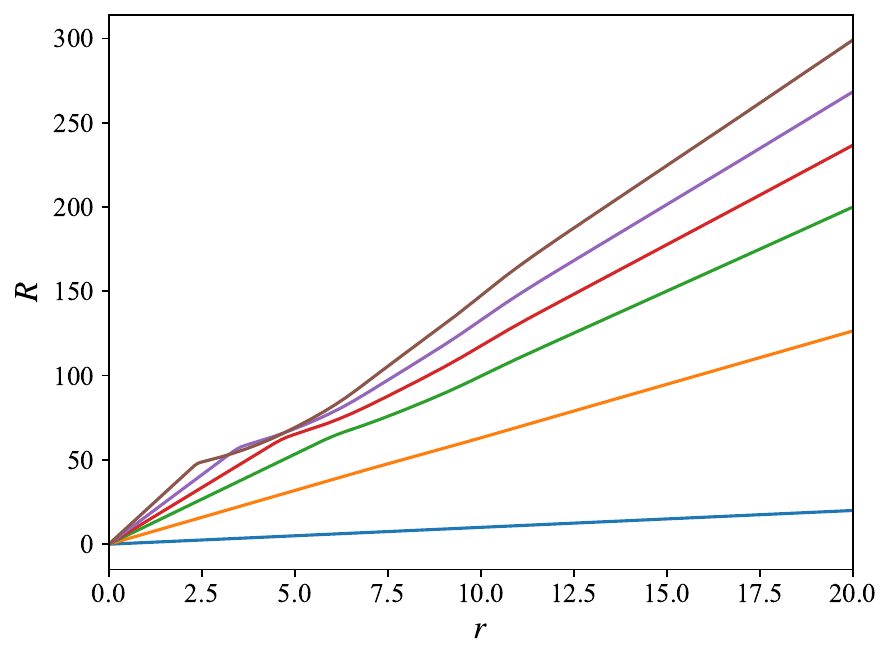}
        \caption{}
    \end{subfigure}
    \\
    \begin{subfigure}{0.495\linewidth}
        \includegraphics[width=\linewidth]{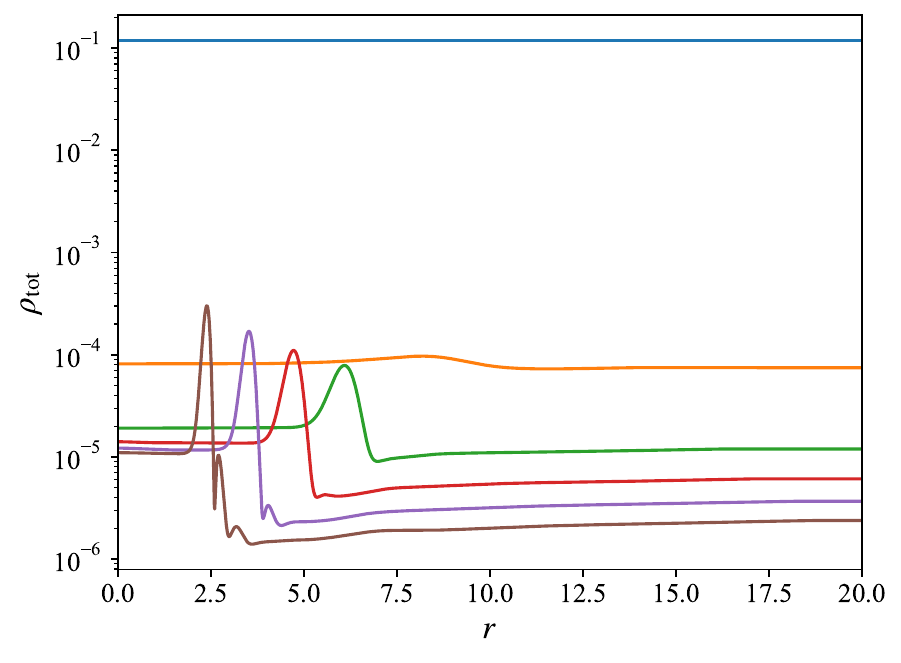}
        \caption{}
    \end{subfigure}
    \begin{subfigure}{0.495\linewidth}
        \includegraphics[width=\linewidth]{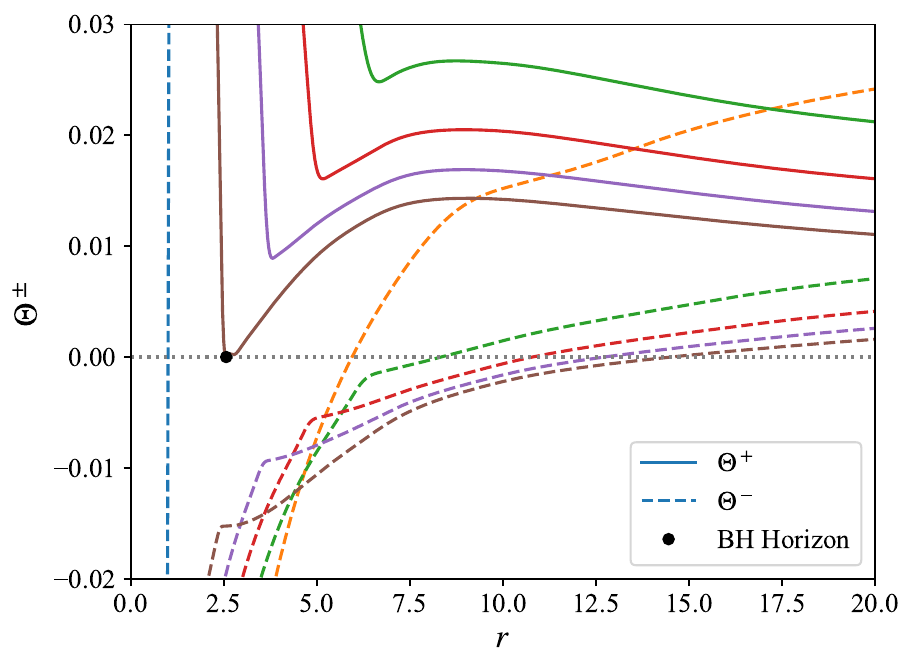}
        \caption{}
    \end{subfigure}
    \\
    \begin{subfigure}{0.495\linewidth}
        \includegraphics[width=\linewidth]{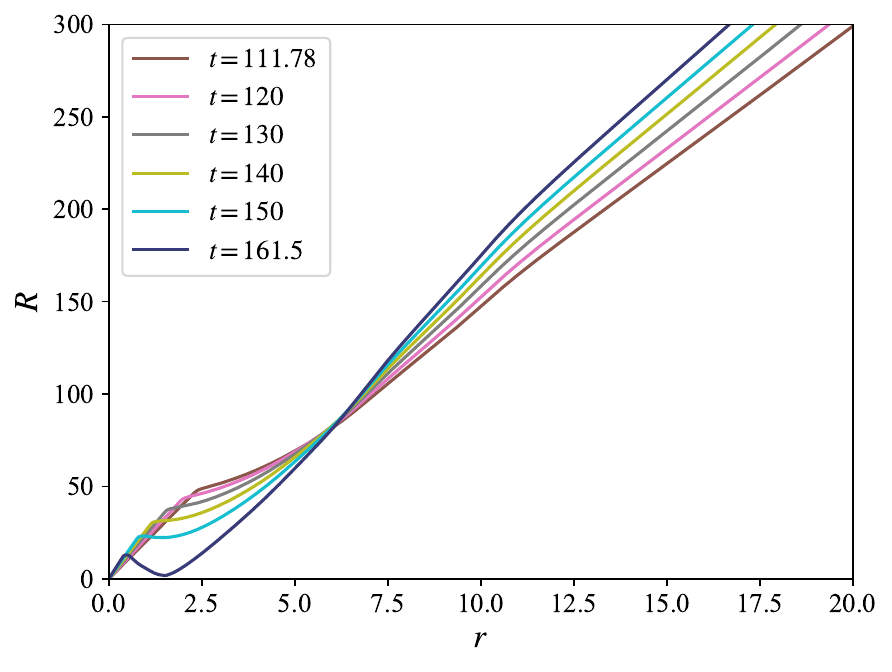}
        \caption{}
    \end{subfigure}
    \begin{subfigure}{0.495\linewidth}
        \includegraphics[width=\linewidth]{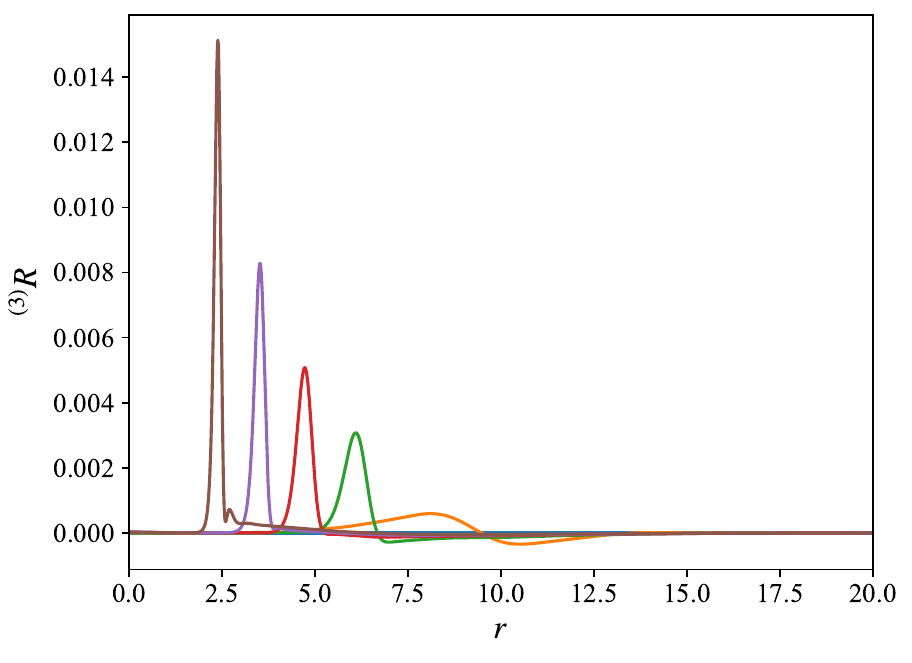}
        \caption{}
    \end{subfigure}
    \caption{Time evolution of several quantities for type A PBH formation with $V_\mathrm{F} = 1\times10^{-5}$. (a) Scalar-field profile $\phi$. (b) Areal radius $R$ before the formation of the apparent horizon. (c) Total energy density $\rho_\mathrm{tot}$. (d) Expansions $\Theta^\pm$; the black dot marks the apparent horizon. (e) Areal radius $R$ after horizon formation. (f) Three-dimensional Ricci scalar ${}^{(3)}R$.}
    \label{fig:typeA}
\end{figure}

If the vacuum energy inside the FVD is insufficient to dominate the dynamics \textit{before} horizon crossing, the bubble wall collapses under self-gravity and can form a type A PBH. In Fig.~\ref{fig:typeA}, we show a representative run with $V_\mathrm{F} = 1\times10^{-5}$.

Panel (a) displays the temporal evolution of the scalar-field profile $\phi$. Similar to the type B case, the bubble wall quickly shrinks under the vacuum-pressure difference. In this case, however, after the wall crosses the horizon (around $t = 50$; see Eq.~\eqref{eq:tH} below), the scalar field begins to oscillate around the true vacuum $\phi = \phi_\mathrm{T}$. The areal radius $R$, shown in panel (b), remains a monotonically increasing function of $r$, although measurable deviations from the FLRW solution persist.

Panel (c) presents the total energy density profiles. As in the type B case, the energy density inside the FVD approaches the false vacuum value $\rho_\mathrm{V} = V_\mathrm{F} = 1\times10^{-5}$, while the exterior radiation density decreases with cosmic expansion. However, here the energy concentrated at the FVD boundary grows as the wall shrinks: the vacuum energy is converted into kinetic and gradient energy concentrated in the wall. The increasing wall energy drives collapse and leads to PBH formation before vacuum-energy domination. This is what was found analytically in Ref.~\cite{Flores:2024lng} and confirmed numerically here. 

%\black{However, we note that the sub-horizon collapse studied here is the continuation of self-consistent super-horizon initial data. It is not a general treatment of initially sub-horizon FVDs, such as in Ref.~\cite{Flores:2024lng}, which would require solving the Hamiltonian and momentum constraints for non-FLRW initial geometries.}

\black{However, the sub-horizon collapse included in this work should not be confused with a general study of initially sub-horizon FVDs. Our initial data are constructed on a super-horizon slice as compensated scalar/vacuum-radiation isocurvature perturbations with FLRW geometry. When such an FVD later becomes sub-horizon, its scalar profile, radiation density, fluid velocity, and local curvature have all been generated dynamically by the preceding evolution and automatically satisfy the Einstein constraints. By contrast, an initially strongly sub-horizon FVD capable of forming a PBH would generally require a large scalar-field energy density already on the initial slice. In that case, the simple compensated FLRW initial data used here may no longer be applicable. One must solve the Hamiltonian and momentum constraints to construct consistent non-FLRW initial data. Such genuinely sub-horizon initial data are beyond the scope of the present work, which will be explored in the future.}

At $t = 111.78$, an apparent horizon forms near (but outside) the bubble wall, satisfying $\Theta^+ = 0$ and $\Theta^- < 0$; its location is marked by the black dot in panel (d). After formation, the apparent horizon rapidly approaches the singularity $R = 0$, as shown in panel (e). Panel (f) displays the three-dimensional Ricci scalar ${}^{(3)}R$ at different times: a positive curvature develops around the contracting wall and increases as wall energy concentrates, facilitating type A PBH formation.

\subsection{No PBH formation}

\begin{figure}[htbp]
    \centering
    \begin{subfigure}{0.495\linewidth}
        \includegraphics[width=\linewidth]{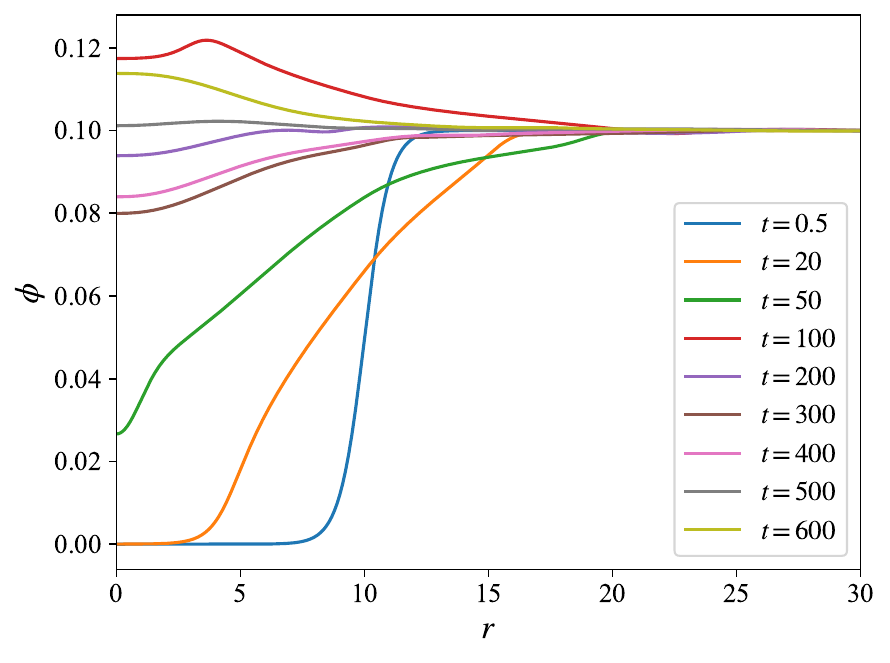}
        \caption{}
    \end{subfigure}
    \begin{subfigure}{0.495\linewidth}
        \includegraphics[width=\linewidth]{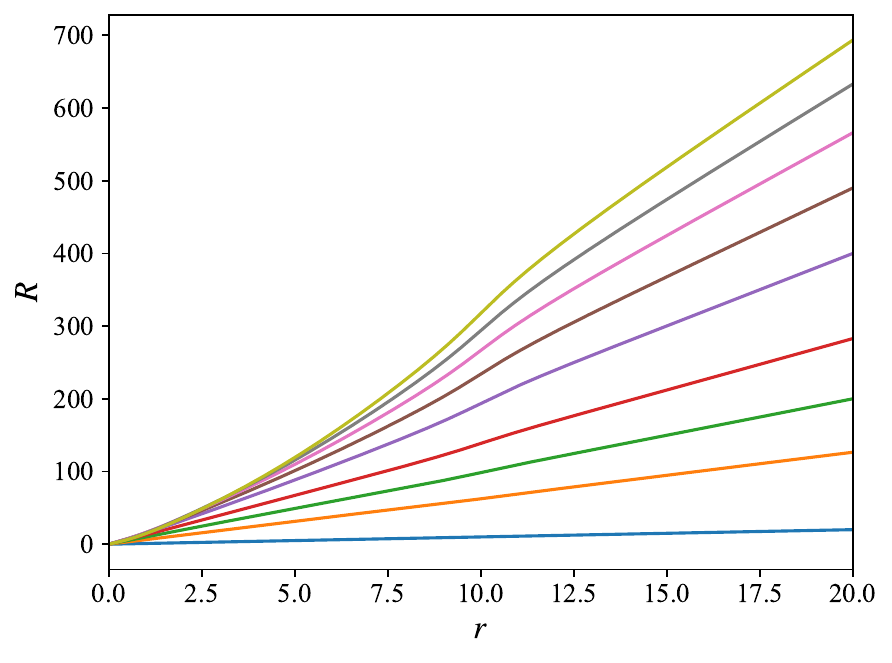}
        \caption{}
    \end{subfigure}
    \\
    \begin{subfigure}{0.495\linewidth}
        \includegraphics[width=\linewidth]{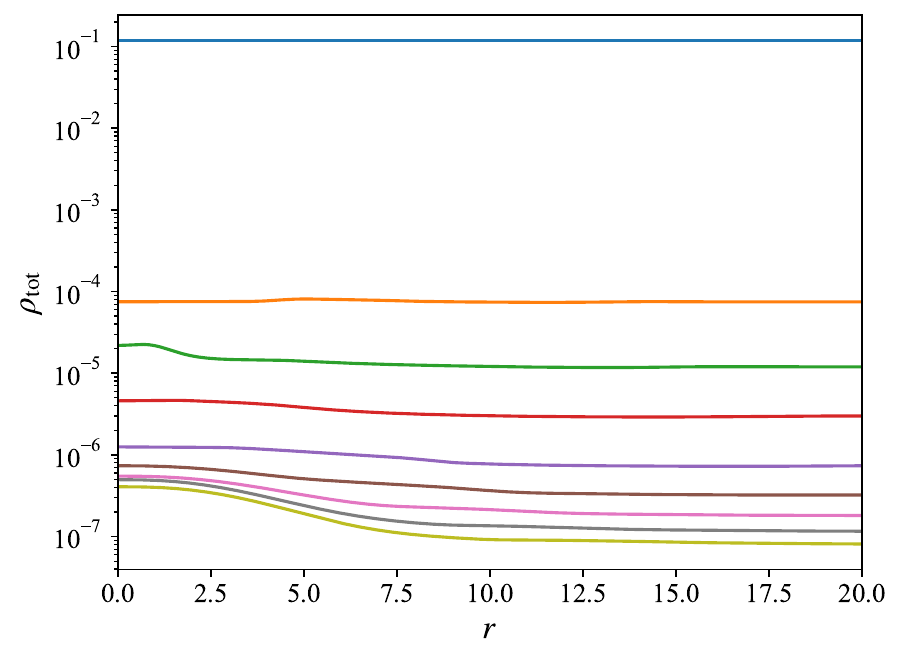}
        \caption{}
    \end{subfigure}
    \begin{subfigure}{0.495\linewidth}
        \includegraphics[width=\linewidth]{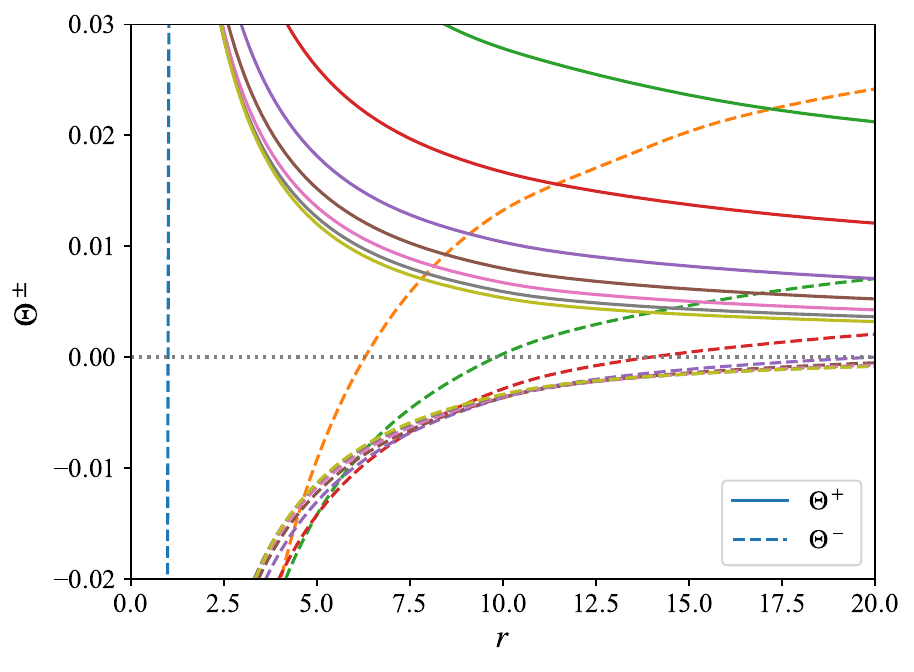}
        \caption{}
    \end{subfigure}
    \\
    \begin{subfigure}{0.495\linewidth}
        \includegraphics[width=\linewidth]{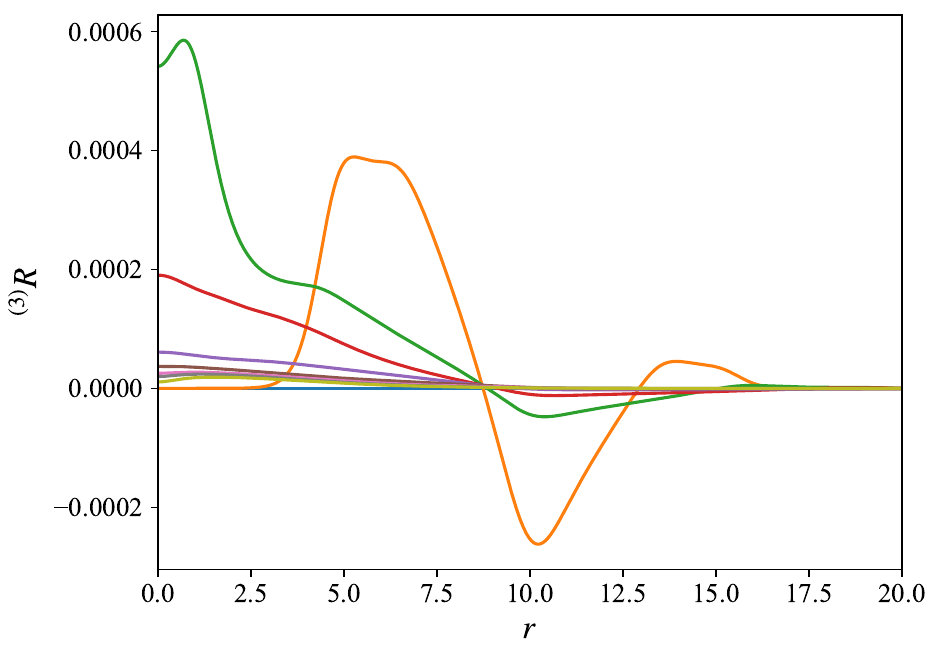}
        \caption{}
    \end{subfigure}
    \caption{Time evolution of several quantities for the no-PBH outcome with $V_\mathrm{F} = 5\times10^{-7}$. (a) Scalar-field profile $\phi$. (b) Areal radius $R$. (c) Total energy density $\rho_\mathrm{tot}$. (d) Expansions $\Theta^\pm$. (e) Three-dimensional Ricci scalar ${}^{(3)}R$.}
    \label{fig:noPBH}
\end{figure}

If the vacuum energy inside the FVD is sufficiently small, the FVD can completely disperse without producing a PBH. Fig.~\ref{fig:noPBH} illustrates this outcome for $V_\mathrm{F} = 5\times10^{-7}$.

Panel (a) shows the scalar-field evolution: the thin-wall profile quickly breaks as the bubble wall contracts. The false-vacuum region vanishes once $\phi(r=0)$ first exceeds $\phi_\mathrm{T}$, which we take as the completion of the phase transition. Thereafter, the scalar field oscillates about the true vacuum $\phi = \phi_\mathrm{T}$~\cite{Cutting:2018tjt, Cutting:2020nla}. During these oscillations, gradient energy remains subdominant while vacuum and kinetic energies exchange periodically. Energy is radiated away as outgoing scalar waves.

The areal-radius profile in panel (b) exhibits only minor departures from the FLRW background. Panel (c) shows that, although the central energy density is still larger than the exterior value, it steadily decays and eventually falls below $V_\mathrm{F} = 5\times10^{-7}$. The expansion profiles in panel (d) confirm the absence of trapped surfaces, and the spatial curvature (panel (e)) relaxes back toward zero. Taken together, these diagnostics indicate that no PBH forms in this parameter regime.

\section{Thresholds for type A/B PBHs} \label{sec:thresholds}

In the previous section, we presented three possible outcomes of FVD evolution: type B PBH formation, type A PBH formation, and no PBH formation, depending on the specific vacuum energy density difference. In this section, we study the thresholds that separate these outcomes by reviewing commonly used PBH-formation criteria and testing them against our simulations.

Both the appearance of an apparent horizon ($\Theta^+ = 0$ and $\Theta^- < 0$) and a bifurcating trapping horizon ($\Theta^+ = \Theta^- = 0$) satisfy
\begin{equation}
    \Theta^+\Theta^- = \frac{2}{R^2}(U^2 - \Gamma^2) = 0 \Longrightarrow \frac{2M}{R} = 1,
\end{equation}
where Eqs.~\eqref{eq:MSmass} and~\eqref{eq:expansion} have been used. Therefore, these horizon-based criteria are equivalent to the Schwarzschild condition $2M/R = 1$ and are independent of the detailed formation mechanism.

Following Refs.~\cite{Deng:2016vzb, Deng:2017uwc, Hashino:2025fse}, we also consider the formation criterion by comparing some characteristic time scales. One of the relevant time scales is the time when the FVD crosses the Hubble horizon. Neglecting cosmological perturbations and assuming a radiation-dominated FLRW background, the horizon crossing time $t_\mathrm{H}$ can be estimated by equating the physical radius of the FVD to the Hubble radius,
\begin{equation} \label{eq:tH}
    R_H(t) = 2t = a(t)r_i = a_i\sqrt{t/t_i}r_i \Longrightarrow t_\mathrm{H} = \frac{r_i^2}{4t_i}.
\end{equation}
This expression can be evaluated from the initial data irrespective of the subsequent dynamics of the FVD. Another relevant time scale is the time when the vacuum energy inside the FVD begins to dominate the local expansion,
\begin{equation}
    t_\mathrm{V} = \frac{1}{2}\sqrt{\frac{3}{8\pi \rho_\mathrm{V}}}.
\end{equation}
We also define a time scale related to the surface tension of the bubble wall,
\begin{equation}
    t_\sigma = \frac{1}{2\pi \sigma}.
\end{equation}
As demonstrated in Ref.~\cite{Maeso:2021xvl, Flores:2024lng, Hashino:2025fse}, $t_\sigma$ is usually much larger than $t_\mathrm{V}$ for sub-Planckian field values,
\begin{equation}
    \frac{t_\sigma}{t_\mathrm{V}} = \frac{\sqrt{8\rho_\mathrm{V}}}{\sqrt{3\pi}\sigma} \simeq \frac{4\sqrt{3}}{\sqrt{\pi}}\frac{\lambda+3}{\lambda^{3/2}\phi_\mathrm{T}} \simeq \frac{4\sqrt{3}}{\sqrt{\pi\lambda}}\frac{1}{\phi_\mathrm{T}} \gg 1 \Longleftrightarrow \phi_\mathrm{T} \ll 4\sqrt{\frac{3}{\pi\lambda}} = 4\sqrt{\frac{24}{\lambda}}M_{\mathrm{Pl}} \simeq 1,
\end{equation}
where we have used the expression~\eqref{eq:sigma} of $\sigma$ in the thin-wall limit and $M_{\mathrm{Pl}} = 1/\sqrt{8\pi G}$ is the reduced Planck mass in our units. Therefore, the surface tension typically has a negligible effect on the PBH formation threshold and will be omitted in what follows. Heuristically, one expects a baby universe/type B PBH outcome if vacuum-energy domination occurs before horizon crossing, $t_\mathrm{H}/t_\mathrm{V} \gtrsim 1$; while a type A PBH or dispersion occurs if horizon crossing precedes vacuum-energy domination, $t_\mathrm{H}/t_\mathrm{V} \lesssim 1$. \black{It is worth noting that the physical mechanism underlying the $t_\mathrm{H}/t_\mathrm{V}$ criterion is closely analogous to PBH formation from cold dark matter (CDM) isocurvature perturbations, as recently studied by Passaglia and Sasaki~\cite{Passaglia:2021jla}. In their scenario, PBH formation requires the CDM isocurvature perturbation to be sufficiently large to induce local matter domination before horizon crossing. Similarly, in our scenario, the initial isocurvature perturbation associated with the FVD must generate a local vacuum-energy domination before or around the horizon-crossing time ($t_\mathrm{H}/t_\mathrm{V} \gtrsim 1$) to successfully produce a type B PBH.}

Another frequently used criterion is the local density contrast at horizon crossing,
\begin{equation}
    \delta(t_\mathrm{H}) = \frac{\rho_\mathrm{in}(t_\mathrm{H}) - \rho_\mathrm{out}(t_\mathrm{H})}{\rho_\mathrm{out}(t_\mathrm{H})},
\end{equation}
where $\rho_\mathrm{in}(t_\mathrm{H})$ and $\rho_\mathrm{out}(t_\mathrm{H})$ are the total energy densities inside and outside the FVD at $t_\mathrm{H}$. In general, if $\delta(t_\mathrm{H})$ exceeds a certain threshold $\delta_c \sim 0.45$, a PBH is expected to form. If we neglect fluid inhomogeneities, the radiation contributions cancel out between inside and outside, and one obtains the simple estimate~\cite{Dent:2025bwo}
\begin{equation}
    \delta(t_\mathrm{H}) \simeq \frac{\rho_\mathrm{V} + \rho(t_\mathrm{H}) - \rho(t_\mathrm{H})}{\rho(t_\mathrm{H})} = \frac{\rho_\mathrm{V}}{\rho(t_\mathrm{H})} = \frac{\rho_\mathrm{V}}{3H^2(t_\mathrm{H})/8\pi} = \frac{8\pi \rho_\mathrm{V}/3}{1/(4t_\mathrm{H}^2)} = \left(\frac{t_\mathrm{H}}{t_\mathrm{V}}\right)^2.
\end{equation}
It follows that the approximate density contrast at $t_\mathrm{H}$ is directly related to the square of the time scale ratio $t_\mathrm{H}/t_\mathrm{V}$. In practice, however, spacetime inhomogeneities and gravitational interactions between the scalar field and the fluid modify this simple estimate, so we compute the true contrast numerically as
\begin{equation}
    \delta(t_\mathrm{H}) = \frac{\rho_\mathrm{tot}(t_\mathrm{H},0) - \rho_\mathrm{tot}(t_\mathrm{H},r_\mathrm{max})}{\rho_\mathrm{tot}(t_\mathrm{H},r_\mathrm{max})}.
\end{equation}

For a given choice of $\phi_\mathrm{T}$, $\lambda$, and $r_i$, we locate the critical value of $V_\mathrm{F}$ that separates type B, type A, and no-PBH outcomes using a bisection search, and we record the corresponding threshold values of $t_\mathrm{H}/t_\mathrm{V}$ and $\delta(t_\mathrm{H})$. This procedure allows us to map how the critical thresholds depend on the physical parameters and to assess which criterion better predicts PBH formation from collapsing FVDs.

\black{We must emphasize that PBH formation in our simulations is not imposed at the horizon-crossing time. The system is evolved from an initially super-horizon FVD through horizon crossing and into the sub-horizon regime, and PBH formation is diagnosed by the actual appearance of an apparent horizon or a bifurcating trapping horizon. Thus, $t_\mathrm{H}/t_\mathrm{V}$ and $\delta(t_\mathrm{H})$ are just convenient, practical, and predictive criteria evaluated at horizon crossing, not statements that the PBH forms exactly at $t_\mathrm{H}$.}

\subsection{Threshold for type B PBH}

\begin{figure}[htbp]
    \centering
    \begin{subfigure}{0.495\linewidth}
        \includegraphics[width=\linewidth]{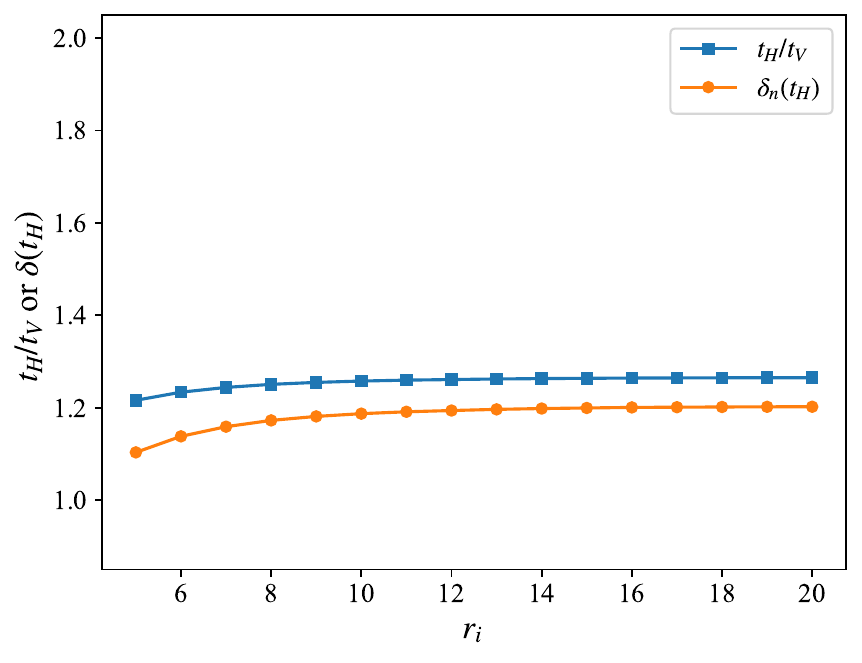}
        \caption{}
    \end{subfigure}
    \begin{subfigure}{0.495\linewidth}
        \includegraphics[width=\linewidth]{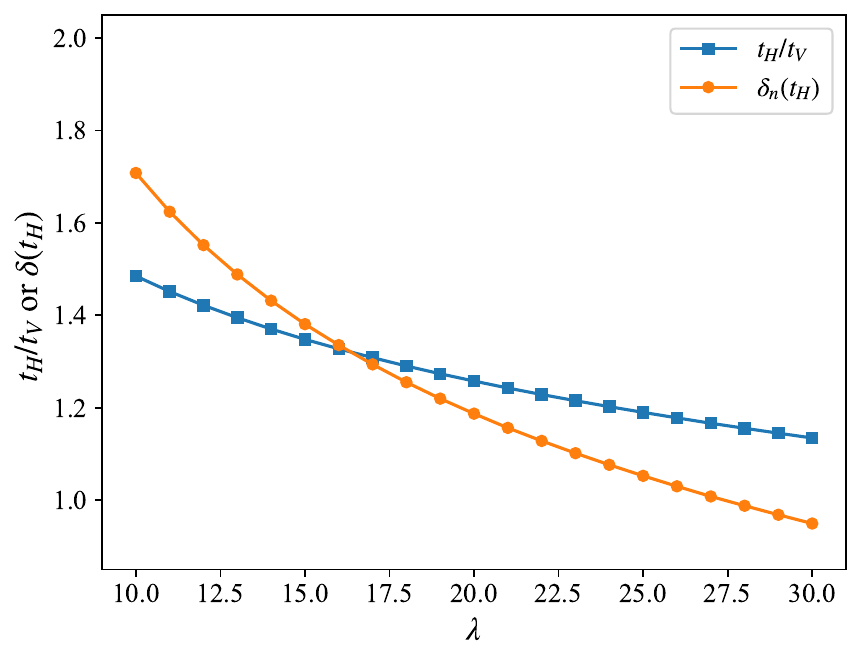}
        \caption{}
    \end{subfigure}
    \\
    \begin{subfigure}{0.495\linewidth}
        \includegraphics[width=\linewidth]{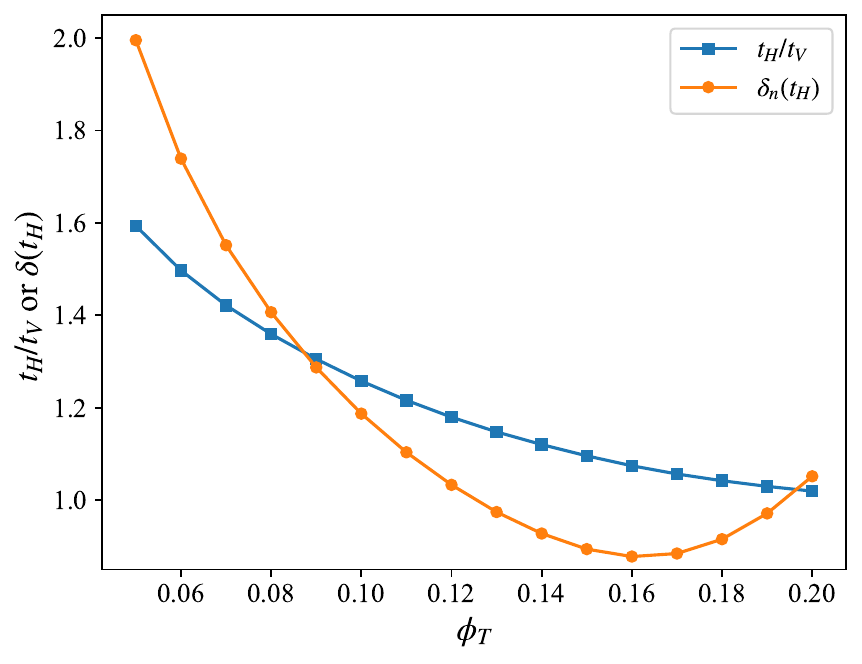}
        \caption{}
    \end{subfigure}
    \caption{Thresholds for type B PBH formation. (a) Dependence on the initial FVD radius $r_i$ (fixed $\phi_\mathrm{T} = 0.1$ and $\lambda = 20$). (b) Dependence on the potential parameter $\lambda$ (fixed $\phi_\mathrm{T} = 0.1$ and $r_i = 10$). (c) Dependence on the true-vacuum field value $\phi_\mathrm{T}$ (fixed $\lambda = 20$ and $r_i = 10$). Blue squares and orange circles denote the critical values of $t_\mathrm{H}/t_\mathrm{V}$ and $\delta(t_\mathrm{H})$, respectively.}
    \label{fig:threshold_typeB}
\end{figure}

We first determine the thresholds for type B PBH formation using the appearance of a bifurcating trapping horizon as the diagnostic. Both the time scale ratio $t_\mathrm{H}/t_\mathrm{V}$ and the density contrast $\delta(t_\mathrm{H})$ refer to the horizon-crossing time $t_\mathrm{H}$ and therefore should be insensitive to $r_i$. To test this, we fix $\phi_\mathrm{T} = 0.1$ and $\lambda = 20$, and then vary $r_i$ to change $t_\mathrm{H}$. The dependence of the thresholds on $r_i$ is shown in panel (a) of Fig.~\ref{fig:threshold_typeB}. It can be seen that the critical values of $t_\mathrm{H}/t_\mathrm{V}$ and $\delta(t_\mathrm{H})$ remain nearly constant as $r_i$ varies, confirming their insensitivity to the initial FVD size.

Next, we study how the thresholds depend on the potential shape. Panel (b) of Fig.~\ref{fig:threshold_typeB} fix $\phi_\mathrm{T} = 0.1$ and $r_i = 10$ and varies $\lambda$. We can see that as $\lambda$ increases (toward the thin-wall regime), both critical $t_\mathrm{H}/t_\mathrm{V}$ and $\delta(t_\mathrm{H})$ decrease. The change in $t_\mathrm{H}/t_\mathrm{V}$ is relatively small: it falls from $\sim 1.5$ to $\sim 1.2$ as $\lambda$ increases from $10$ to $30$, approaching the thin-wall estimate $t_\mathrm{H}/t_\mathrm{V} \sim 1.25$ reported in Ref.~\cite{Hashino:2025fse}. The critical $\delta(t_\mathrm{H})$ shows a larger decline with increasing $\lambda$, consistent with the trend noted in that work.

\begin{figure}[htbp]
    \centering
    \includegraphics[width=0.495\linewidth]{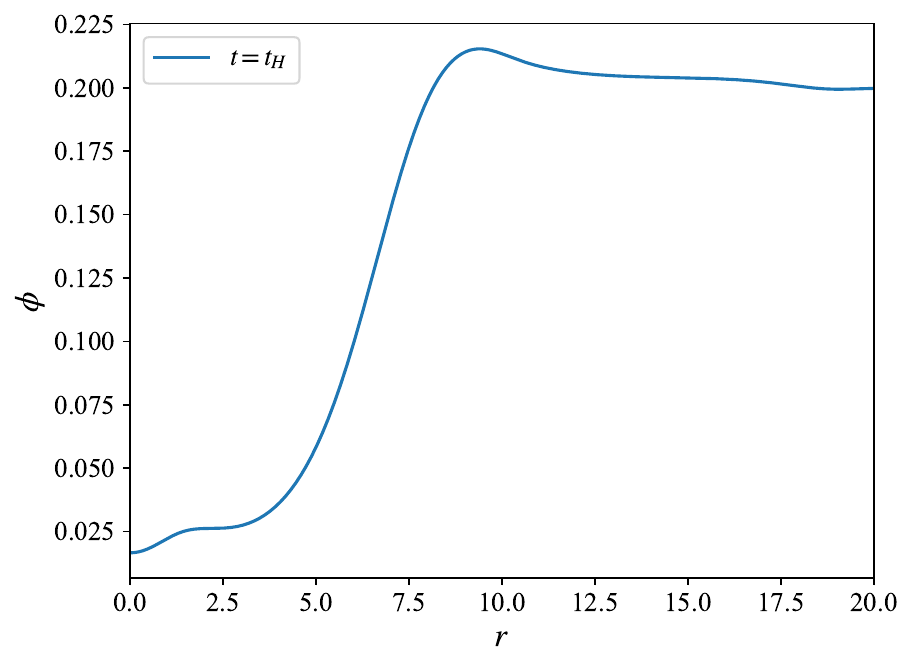}
    \includegraphics[width=0.495\linewidth]{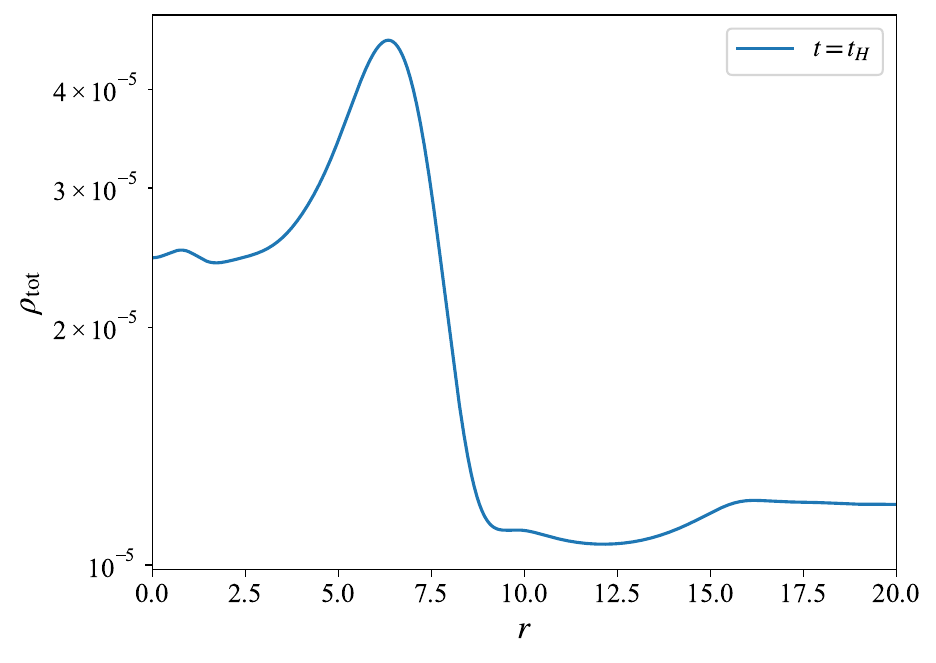}
    \caption{Scalar field (left) and total energy density (right) profiles at the horizon crossing time for $\phi_\mathrm{T} = 0.2$ with $V_\mathrm{F}$ near the critical threshold. The thin-wall profile is distorted, and the central field value $\phi(r=0)$ departs from the false-vacuum value.}
    \label{fig:hc}
\end{figure}

Finally, we examine the sensitivity of the thresholds on the true-vacuum field value $\phi_\mathrm{T}$. Panel (c) of Fig.~\ref{fig:threshold_typeB} displays the dependence on $\phi_\mathrm{T}$ (with $\lambda = 20$ and $r_i = 10$ fixed). We find that the critical $t_\mathrm{H}/t_\mathrm{V}$ decreases modestly as $\phi_\mathrm{T}$ grows (roughly from $1.6$ to $1.1$ as $\phi_\mathrm{T}$ increases from $0.05$ to $0.2$), while $\delta(t_\mathrm{H})$ exhibits a non-monotonic behavior: it first decreases and then increases. This nonmonotonicity likely arises because a large $\phi_\mathrm{T}$ disrupts the thin-wall configuration, degrading the definition of the FVD boundary and thereby affecting the computation of the local contrast. Fig.~\ref{fig:hc} illustrates this effect for $\phi_\mathrm{T} = 0.2$ with $V_\mathrm{F}$ near the threshold, showing the scalar field and total energy density profiles at the horizon crossing time. It can be seen that the thin-wall profile is distorted and the central field value $\phi(r=0)$ departs from the false-vacuum value.

Across the parameter ranges we explored, the critical $t_\mathrm{H}/t_\mathrm{V}$ varies relatively small and remains between $1.1$ and $1.6$, indicating that the time-scale criterion is robust to changes in potential parameters and initial FVD size, consistent with the results obtained in the thin-wall limit~\cite{Hashino:2025fse}. In contrast, the local density contrast $\delta(t_\mathrm{H})$ is more sensitive to model details and thus less reliable as a universal threshold for type B PBH formation in FVD collapse. Our fully nonlinear treatment yields larger variations in the critical values of $\delta(t_\mathrm{H})$ than those reported in thin-wall analyses, reflecting breakdowns of the thin-wall approximation in some parameter regions and complex nonlinear dynamics during collapse.

\subsection{Threshold for type A PBH}

\begin{figure}[htbp]
    \centering
    \begin{subfigure}{0.495\linewidth}
        \includegraphics[width=\linewidth]{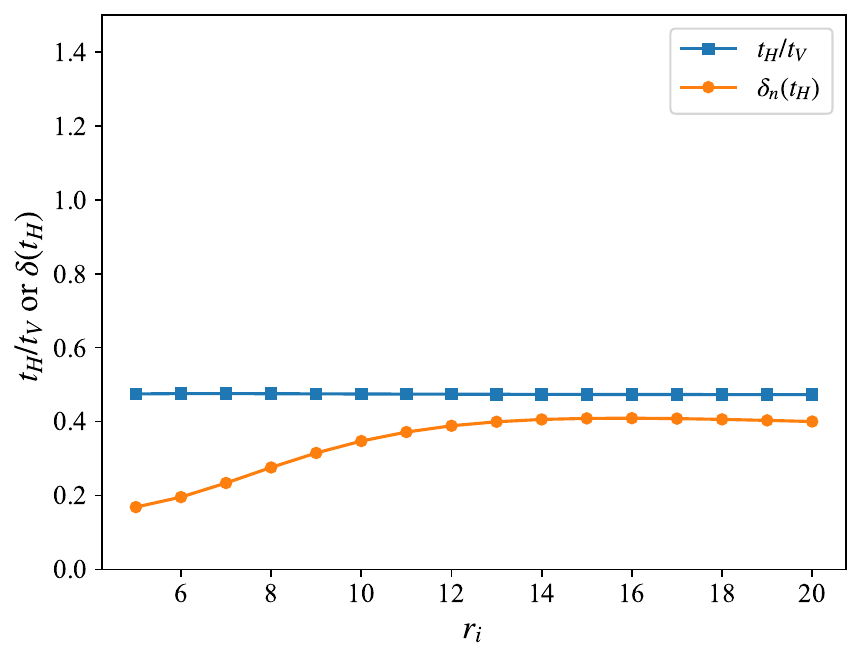}
        \caption{}
    \end{subfigure}
    \begin{subfigure}{0.495\linewidth}
        \includegraphics[width=\linewidth]{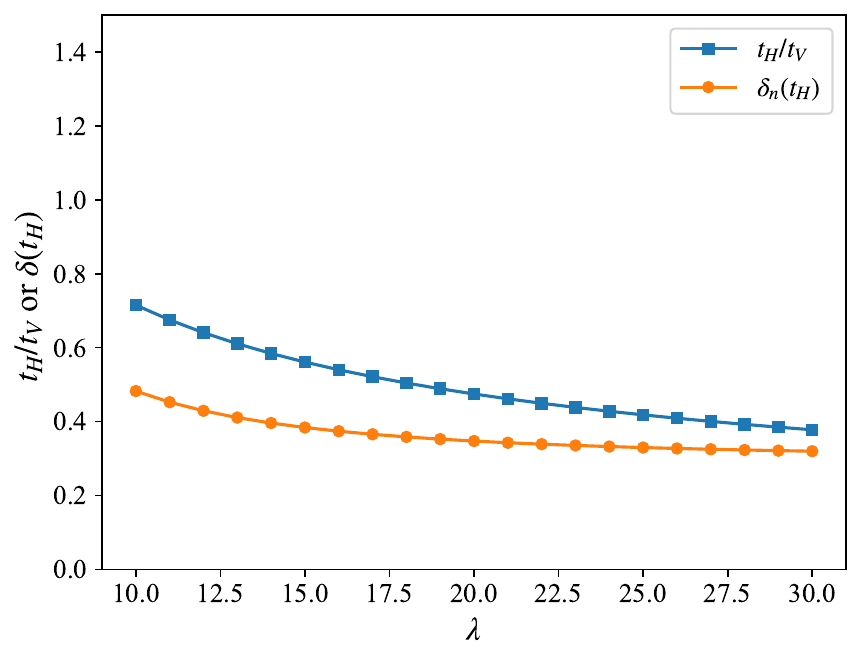}
        \caption{}
    \end{subfigure}
    \\
    \begin{subfigure}{0.495\linewidth}
        \includegraphics[width=\linewidth]{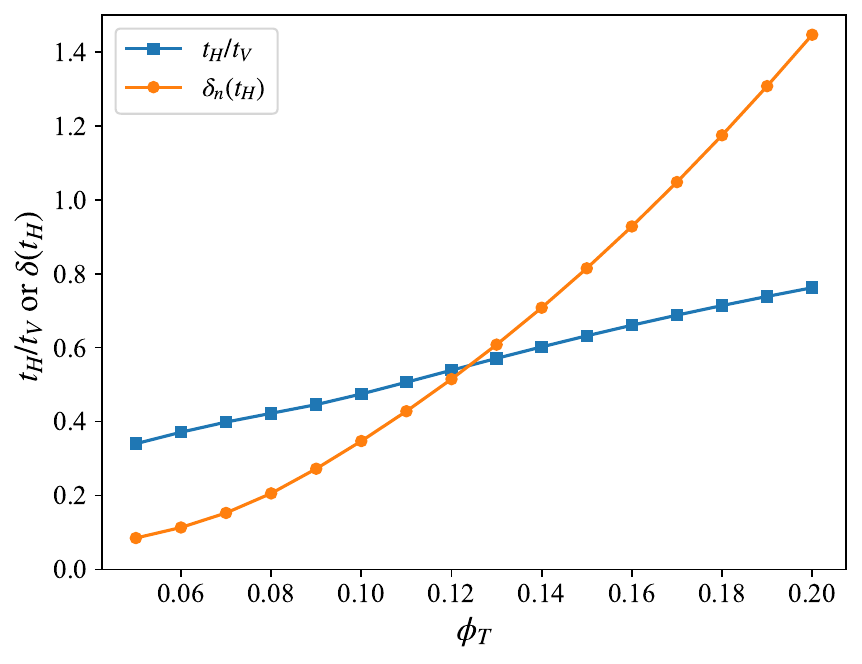}
        \caption{}
    \end{subfigure}
    \caption{Thresholds for type A PBH formation. (a) Dependence on the initial FVD radius $r_i$ (fixed $\phi_\mathrm{T} = 0.1$ and $\lambda = 20$). (b) Dependence on the potential parameter $\lambda$ (fixed $\phi_\mathrm{T} = 0.1$ and $r_i = 10$). (c) Dependence on the true-vacuum field value $\phi_\mathrm{T}$ (fixed $\lambda = 20$ and $r_i = 10$). Blue squares and orange circles denote the critical values of $t_\mathrm{H}/t_\mathrm{V}$ and $\delta(t_\mathrm{H})$, respectively.}
    \label{fig:threshold_typeA}
\end{figure}

For smaller vacuum energy, the scalar field value at the center $\phi(r=0)$ can first cross to the true vacuum and then oscillate about $\phi_\mathrm{T}$. During these oscillations, the vacuum energy may be converted into kinetic and gradient energy that can still drive collapse and produce a type A PBH. In practice, the smallest black hole we reliably resolve in our simulations has comoving radius $r_{\rm BH} \approx 0.5$. For parameters closer to the threshold, the central field profile becomes extremely sharp, and our code crashes before horizon formation is unambiguously detected; we therefore treat the last resolvable run as the approximate threshold between type A PBH formation and dispersion.

Fig.~\ref{fig:threshold_typeA} summarizes how the type A thresholds depend on $r_i$, $\lambda$, and $\phi_\mathrm{T}$. As in the type B case, the critical values of $t_\mathrm{H}/t_\mathrm{V}$ are largely insensitive to $r_i$ (panel (a)), indicating that it is independent of the horizon crossing time. By contrast, $\delta(t_\mathrm{H})$ exhibits large variation with $r_i$ since a small $r_i$ often destroys the thin-wall configuration before horizon crossing, making the FVD boundary ill-defined and the density contrast measurement uncertain.

Varying $\phi_\mathrm{T}$ produces a similar effect (panel (c)): $t_\mathrm{H}/t_\mathrm{V}$ changes only slightly (roughly from $0.35$ to $0.7$) as $\phi_\mathrm{T}$ increases from $0.05$ to $0.2$, while $\delta(t_\mathrm{H})$ can grow dramatically, maybe because the bubble wall energy is sensitive to $\phi_\mathrm{T}$ and large $\phi_\mathrm{T}$ tends to distort the thin-wall profile before horizon crossing. When $\lambda$ is varied from $10$ to $30$ (panel (b)), $\delta(t_\mathrm{H})$ is comparatively stable and provides a useful criterion, typically lying in the range $\sim 0.35 - 0.5$, close to the conventional threshold for PBH formation from curvature perturbations in radiation domination.

In summary, the time scale ratio $t_\mathrm{H}/t_\mathrm{V}$ serves as a PBH-formation criterion that is relatively independent of the thin-wall assumption, while the local density contrast $\delta(t_\mathrm{H})$ depends strongly on whether a well-defined thin wall (and hence a clear FVD boundary) exists. If the thin-wall configuration is destroyed before horizon crossing, the FVD boundary becomes ambiguous, leading to significant uncertainty in $\delta(t_\mathrm{H})$. In our survey, the critical value of $t_\mathrm{H}/t_\mathrm{V}$ for type B PBH formation lies roughly in the range $1.1 - 1.6$, while for type A PBH formation it lies roughly in the range $0.35 - 0.7$. For cases where the thin-wall approximation holds, the critical value of $\delta(t_\mathrm{H})$ for type B PBH formation spans approximately $1 - 1.7$, while for type A PBH formation it remains relatively stable between $0.35$ and $0.5$.

\black{The sensitivity of $\delta(t_\mathrm{H})$ to the wall structure is reminiscent of the dependence of the collapse threshold on the perturbation profile in the context of PBH formation from adiabatic curvature perturbations~\cite{Musco:2018rwt, Musco:2020jjb, Pi:2024ert}. In this sense, the density profile generated by the collapsing FVD is controlled by the physical properties of the domain and its wall, such as the initial FVD radius, the scalar potential parameters, and the vacuum-energy difference. Our parameter scan over $r_i$, $\lambda$, and $\phi_\mathrm{T}$ therefore partly probes the profile dependence of FVD collapse. The results show that $\delta(t_\mathrm{H})$ is sensitive to these details, especially when the thin-wall configuration breaks down and the FVD boundary becomes ambiguous. This behavior is consistent with the known profile dependence of PBH thresholds. By contrast, the timescale ratio $t_\mathrm{H}/t_\mathrm{V}$ is less sensitive to the detailed wall profile and captures the main competition between vacuum-energy domination and horizon crossing. We therefore regard $t_\mathrm{H}/t_\mathrm{V}$ as a more robust criterion for the class of FVD configurations studied here, while a full treatment of the statistical distribution of FVD sizes and profiles from bubble nucleation is left for future work.}

\section{Conclusions and discussions} \label{sec:conclusions}

In this work, we have performed fully nonlinear, spherically symmetric numerical simulations of FVD collapse in a coupled gravity-scalar-fluid system to study the dynamical pathways to PBH formation during delayed FOPTs and to test PBH-formation criteria. We adopted a phenomenological scalar potential with adjustable parameters to model the FOPTs, and set up initial conditions corresponding to superhorizon FVDs embedded in a radiation-dominated FLRW background. Numerically, we evolved the coupled system with the method of lines and a Berger-Oliger AMR scheme to resolve the bubble wall.

Our simulations reveal three distinct dynamical outcomes, depending primarily on the competition between vacuum energy inside the FVD and gravitational collapse: 
\begin{enumerate}
    \item type B (supercritical) PBH formation, in which the FVD interior inflates, a wormhole throat appears, and a bifurcating trapping horizon ($\Theta^+ = \Theta^- = 0$) forms, producing a baby-universe interior that is observed externally as a PBH; 
    \item type A (subcritical) PBH formation, where the wall energy concentrates, causing the wall to collapse under self-gravity to form an apparent horizon ($\Theta^+ = 0$ and $\Theta^- < 0$) without an interior baby universe; 
    \item no-PBH outcome, in which the false vacuum region disperses, the scalar field oscillates about the true vacuum, energy is radiated away as outgoing scalar waves, and no trapped surface forms.
\end{enumerate}
We traced these behaviors through the scalar-field profile, areal-radius evolution, total energy-density, expansions, and the spatial Ricci scalar.

A primary goal of this work is to assess practical criteria for PBH formation arising from the collapse of FVDs. We focus on two diagnostics: the time-scale ratio $t_\mathrm{H}/t_\mathrm{V}$ (horizon-crossing time versus vacuum-energy domination time) and the local density contrast $\delta(t_\mathrm{H})$ evaluated at horizon crossing. Our main findings are:

\begin{itemize}

\item The horizon-based diagnostics (apparent horizon or bifurcating trapping horizon) are equivalent to the Schwarzschild condition $2M/R = 1$ in our coordinates and provide a robust, mechanism-independent indicator for PBH formation.

\item For the sub-Planckian parameter ranges we considered, surface-tension effects (characterized by $t_\sigma$) are typically subdominant and do not qualitatively affect the PBH-formation thresholds.

\item The time-scale ratio $t_\mathrm{H}/t_\mathrm{V}$ is comparatively stable across parameter variations. For the parameter ranges we explored ($\phi_\mathrm{T}$, $\lambda$, $r_i$ varied around our representative thin-wall choice), the critical $t_\mathrm{H}/t_\mathrm{V}$ separating type B formation lies roughly in the range $\sim 1.1 - 1.6$ (approaching the thin-wall estimation $\sim 1.25$~\cite{Hashino:2025fse} as $\lambda$ increases), while for type A formation, it lies approximately in $\sim 0.35-0.7$.

\item By contrast, the local density contrast $\delta(t_\mathrm{H})$ depends sensitively on microphysical and geometrical details. When a clear thin-wall FVD boundary exists, $\delta(t_\mathrm{H})$ can provide a useful threshold (typical values for type A lie near $\sim 0.35-0.5$); however, when the thin-wall approximation breaks down (e.g. for small $r_i$ or large $\phi_\mathrm{T}$), the FVD boundary becomes ill-defined and $\delta(t_\mathrm{H})$ varies widely, limiting its reliability as a universal criterion.

\end{itemize}

These results have several implications. First, they demonstrate that delayed FOPTs can produce PBHs through qualitatively different channels (baby-universe formation versus direct collapse). Second, a simple time-scale comparison $t_\mathrm{H}/t_\mathrm{V}$ captures the essential competition between expansion driven by vacuum energy and gravitational collapse, and can serve as a robust predictor for PBH formation across a wide range of parameters. Third, adopting a single density threshold (e.g. $\delta_c \sim 0.45$ derived from curvature-perturbation-driven collapse) is not generally justified for FVD-driven PBH formation without careful assessment of wall structure and nonlinear dynamics.

We must emphasize several caveats and limitations of the present study. (i) Spherical symmetry was assumed throughout the paper. This can be ensured for inflationary bubbles, but for delayed FOPTs, FVD remnants may have complex, non-spherical, or even arbitrary shapes, which is hard to extract definite results. \black{This may modify, and in many cases is expected to suppress, the collapse of FVDs that would have formed PBHs in spherical symmetry.} We also neglected possible nucleation of additional bubbles inside the FVD, which would further complicate the dynamics. These effects could modify collapse dynamics and thresholds. (ii) We adopted a temperature-independent, phenomenological scalar potential and assumed only gravitational coupling between the scalar and radiation fluid. Nontrivial finite-temperature effects and scalar-fluid couplings can change wall speed, wall permeability, and collapse dynamics. \black{Moreover, our radiation component is modeled as a perfect fluid, but dissipative effects such as viscosity, heat conduction, finite-mean-free-path diffusion, or Silk damping~\cite{Silk:1967kq, Domenech:2025bvr}, may suppress the growth of compactness.} (iii) We use geodesic slicing $A = 1$ for numerical stability. However, some diagnostics---especially $\delta(t_\mathrm{H})$---are gauge-dependent. Our results echo recent studies and underscore the need for gauge-invariant measures where possible. (iv) Finally, we did not compute PBH mass functions or abundances integrated over stochastic nucleation statistics; such predictions require combining our dynamical thresholds with a probabilistic model of FVD occurrence. \black{In this work, we have focused on the local nonlinear evolution of a single isolated, approximately spherical FVD. The thresholds obtained here should therefore be interpreted as conditional collapse criteria for a given FVD configuration. To obtain the abundance, these criteria must be combined with the probability distribution of FVD sizes and profiles generated by the underlying bubble nucleation history. In the simplest treatment, bubble nucleation is often approximated as a Poisson process. However, critical bubbles in FOPTs may be biased and spatially correlated~\cite{Pirvu:2021roq, DeLuca:2021mlh}. If present, bubble correlations can modify the probability of forming large FVDs, their spatial distribution, and their departure from spherical symmetry. Incorporating these effects into the PBH abundance calculation is beyond the scope of the present work and is left for future study.}

Looking forward, several directions are natural extensions of this work. \black{First, large inhomogeneities generated by delayed FOPTs may also induce curvature perturbations that source second-order scalar-induced gravitational waves~\cite{Liu:2022lvz, Elor:2023xbz, Cai:2024nln, Lewicki:2024ghw, Franciolini:2025ztf, Wang:2026zvz}. Together with PBH production, such a stochastic gravitational wave background can serve as a complementary probe for delayed FOPTs. Second, direct (non-gravitational) couplings between the scalar field and additional species or the surrounding plasma could lead to particle production~\cite{Shakya:2023kjf}. Such non-perturbative particle production may act as an effective friction or dissipation mechanism, transferring energy from the moving wall/vacuum sector to radiation or other species. This could potentially alter the collapse dynamics and the PBH formation thresholds. These effects are model-dependent and should be incorporated in future simulations by adding appropriate matter components or source terms. It is also important to include temperature-dependent potentials to assess robustness against finite-temperature effects. Third, PBH spin is absent in the present spherically symmetric setup. Removing the symmetry restriction by performing fully $(3+1)$-dimensional numerical-relativity simulations will enable us to test non-spherical collapse channels and study the spin of PBHs formed from delayed FOPTs~\cite{Harada:2020pzb, Banerjee:2023qya}. Finally, a complementary avenue is to embed these dynamical results into statistical models of bubble nucleation to compute PBH mass spectra and cosmic abundances. FVDs originate from the stochastic and asynchronous nature of bubble nucleation, and their distribution is generally expected to be non-Gaussian. Our results should be viewed as conditional collapse criteria for a given FVD configuration. A complete PBH abundance calculation must convolve these thresholds with the statistical distribution of FVD sizes and profiles generated by the underlying FOPT.}

In summary, our fully nonlinear simulations clarify the dynamical pathways by which delayed FOPTs can produce PBHs and quantify practical diagnostics for distinguishing different dynamical outcomes. The time-scale ratio $t_\mathrm{H}/t_\mathrm{V}$ emerges as a robust, physically transparent predictor across a wide region of parameter space, while the local density contrast $\delta(t_\mathrm{H})$ requires careful application. These results provide a foundation for more realistic, multidimensional studies that aim to link phase-transition microphysics to observable PBH signatures.

\appendix
\section{Derivations of $K$ and ${}^{(3)}R$} \label{app:K}

In this appendix, we derive the expressions for the trace of the extrinsic curvature $K$ and the three-dimensional Ricci scalar ${}^{(3)}R$ used in the main text. Greek and Latin letters denote spacetime and spatial indices, respectively. A general $3+1$ line element can be written as
\begin{equation}
    \mathrm{d}s^2 = -A^2\mathrm{d}t^2 + \gamma_{ij}(\mathrm{d}x^i + \beta^i\mathrm{d}t)(\mathrm{d}x^j + \beta^j\mathrm{d}t),
\end{equation}
where $A$ is the lapse function, $\beta^i$ is the shift vector, and $\gamma_{ij}$ is the spatial metric. The unit normal vector to the spatial hypersurfaces is
\begin{equation}
    n^\mu = (1/A,-\beta^i/A), \quad n_\mu = (-A,\boldsymbol{0}),
\end{equation}
and the spatial projection operator is $\gamma^\mu_{\nu} \equiv \delta^\mu_{\nu} + n^\mu n_\nu$. The extrinsic curvature is defined by
\begin{equation}
    K_{\mu\nu} \equiv -\gamma^\alpha_{\mu}\nabla_\alpha n_\nu = -\frac{1}{2}\mathcal{L}_n \gamma_{\mu\nu},
\end{equation}
where $\nabla_\alpha$ is the covariant derivative associated with the spacetime metric $g_{\mu\nu}$, and $\mathcal{L}_n$ denotes the Lie derivative along $n^\mu$. Its trace is
\begin{equation}
    K \equiv g^{\mu\nu}K_{\mu\nu} = \gamma^{ij}K_{ij}.
\end{equation}
The time derivative of $K$ is given by (see, e.g., Refs.~\cite{2008itnr.book.....A, 2016nure.book.....S})
\begin{equation} \label{eq:pt_K}
    \partial_t K = \beta^i \partial_i K - \gamma^{ij} D_i D_j A + A \left( K_{ij} K^{ij} + 4\pi (E + S) \right),
\end{equation}
where $D_i$ is the covariant derivative associated with the spatial metric $\gamma_{ij}$, $E \equiv n^\mu n^\nu T_{\mu\nu}$, and $S \equiv g^{\mu\nu} \gamma^\alpha_{\mu} \gamma^\beta_{\nu} T_{\alpha\beta} = \gamma^{ij} T_{ij}$. The three-dimensional Ricci scalar ${}^{(3)}R$ of the spatial hypersurfaces is
\begin{equation}
    {}^{(3)}R = \gamma^{ij}\left(\partial_k \Gamma^k_{ij} - \partial_j \Gamma^k_{ik} + \Gamma^l_{ij}\Gamma^k_{kl} - \Gamma^l_{ik}\Gamma^k_{jl}\right),
\end{equation}
where $\Gamma^k_{ij}$ is the Christoffel symbol associated with $\gamma_{ij}$.

Specializing to the comoving threading $\beta^i = 0$ and spherical symmetry, the line element reduces to Eq.~\eqref{eq:metric}, and the spatial metric becomes
\begin{equation}
    \gamma_{ij} = \mathrm{diag}\left(B^2, R^2, R^2\sin^2\theta\right).
\end{equation}
The nonzero components of the extrinsic curvature are
\begin{equation}
    K_{ij} = \mathrm{diag}\left(-\frac{B\dot{B}}{A}, -\frac{R\dot{R}}{A}, -\frac{R\dot{R}\sin^2\theta}{A}\right),
\end{equation}
from which the trace becomes
\begin{equation} \label{eq:K_app}
    K = -\frac{1}{A}\left(\frac{\dot{B}}{B} + \frac{2\dot{R}}{R}\right) = -\frac{1}{A}\left(\frac{\dot{B}}{B} + \frac{2AU}{R}\right),
\end{equation}
and a straightforward calculation yields
\begin{equation}
    K_{ij} K^{ij} = \frac{1}{A^2}\left[\left(\frac{\dot{B}}{B}\right)^2 + 2\left(\frac{\dot{R}}{R}\right)^2\right] = \left(K+\frac{2U}{R}\right)^{2}+2\left(\frac{U}{R}\right)^{2},
\end{equation}
where in the last equality we have used $\dot{B} = -AB(K+2U/R)$ from Eq.~\eqref{eq:K_app}. Substituting this expression together with the comoving threading $\beta^i = 0$ and geodesic slicing $A = 1$ into Eq.~\eqref{eq:pt_K} reproduces Eq.~\eqref{eq:K} in the main text. Finally, the three-dimensional Ricci scalar reduces to the form
\begin{equation}
    {}^{(3)}R = \frac{2}{R^2}\left[1 - \frac{(R')^2}{B^2} - \frac{2R}{B}\left(\frac{R'}{B}\right)'\right].
\end{equation}

\section{Convergence tests} \label{app:convergence}

In this appendix, we validate our numerical implementation by performing a convergence test similar to Refs.~\cite{Ning:2025ogq, Ning:2023edr}. To quantify the numerical error, we use the $L_2$-norm of Eq.~\eqref{eq:error}, which is actually a combination of the Hamiltonian and momentum constraints and should vanish if the Einstein-scalar-fluid system is solved correctly. Concretely, we define the constraint violation and its $L_2$-norm as
\begin{subequations}
  \begin{align}
    \mathcal{H} &\equiv M' - 4\pi R^{2}\left(R'T_{00}-\dot{R}T_{01}\right),\\
    {||\mathcal{H}||}_2 &\equiv \frac{1}{N}\sqrt{\sum_{i=1}^N{|\mathcal{H}_i|}^2},
  \end{align}
\end{subequations}
where the subscript $i$ labels grid points and $N$ is the number of grid points.

\begin{figure}[htbp]
  \centering
  \includegraphics[width=0.495\linewidth]{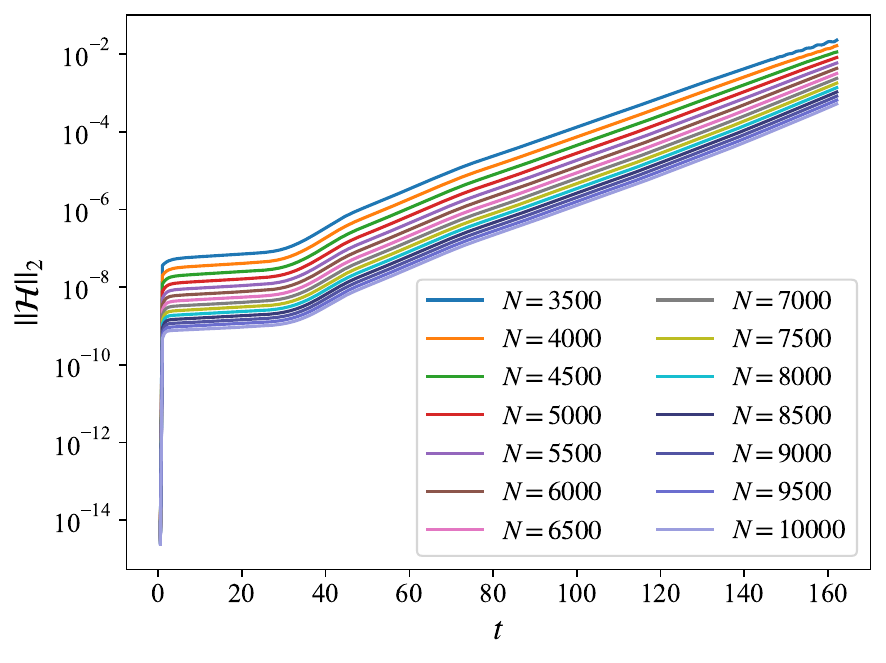}
  \includegraphics[width=0.495\linewidth]{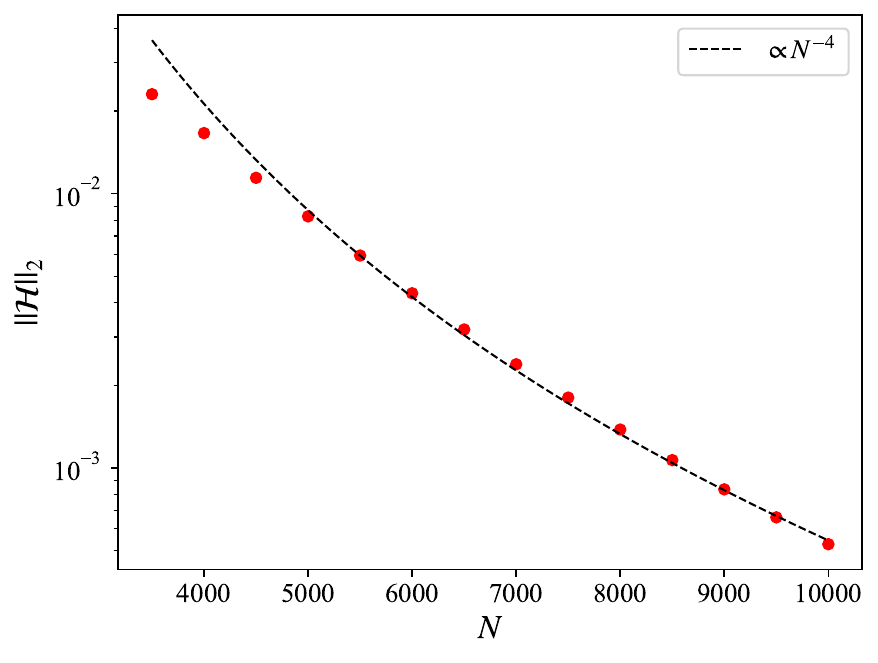}
  \caption{Convergence test of our numerical code. \textit{Left panel}: Time evolution of ${||\mathcal{H}||}_2$ for several grid resolutions. \textit{Right panel}: the $L_2$-norm at the end of the simulation as a function of grid point number $N$. The black dashed line indicates the expected $N^{-4}$ scaling.}
  \label{fig:convergence}
\end{figure}

To probe the convergence order of our numerical code, we disable AMR and run a sequence of fixed-resolution simulations. We choose the type B PBH formation case ($V_\mathrm{F} = 5\times10^{-5}$) for this test, since it involves the most complex dynamics among the three outcomes. Grid resolutions are varied from $N = 3500$ to $N = 10000$ in increments of $500$, while all other parameters remain as in the main text. The left panel of Fig.~\ref{fig:convergence} shows the time evolution of ${||\mathcal{H}||}_2$ for selected grid resolutions. It can be seen that for sufficiently high resolution, the constraint violation remains at the level $\mathcal{O}(10^{-3})$ throughout the simulation. The right panel plots the $L_2$-norm at the end of the simulation as a function of $N$. A reference $N^{-4}$ scaling (black dashed line) matches the data well, confirming fourth-order convergence consistent with our fourth-order finite-difference spatial discretization and fourth-order Runge-Kutta time integrator.

\acknowledgments

We thank Zi-Yan Yuwen for insightful discussions. This work is supported by the National Key Research and Development Program of China Grant No. 2021YFC2203004, No. 2021YFA0718304, and No. 2020YFC2201501, the National Natural Science Foundation of China Grants No. 12422502, No. 12547110, No.12588101, No. 12235019, and No. 12447101, and the Science Research Grants from the China Manned Space Project with No. CMS-CSST-2025-A01.

% \paragraph{Note added.} This is also a good position for notes added
% after the paper has been written.

% Bibliography

\bibliographystyle{JHEP}
\bibliography{biblio.bib}

\end{document}